\documentclass[12pt]{amsart}

\usepackage[leqno]{amsmath}
\usepackage{amssymb}
\usepackage[normalem]{ulem}
\usepackage[]{graphicx}
\usepackage[margin=3cm]{geometry}
\usepackage[]{color}

\newtheorem{thm}{Theorem}[section]
\newtheorem{lem}[thm]{Lemma}

\newtheorem{cor}[thm]{Corollary}
\newtheorem{rem}[thm]{Remark}
\theoremstyle{definition}

\newcommand{\N}{\mathcal N}
\newcommand{\bN}{\mathbf N}
\newcommand{\T}{\mathbf T}
\newcommand{\TT}{\mathcal T}

\newcommand{\B}{\mathbb B}

\newcommand{\C}{\mathcal C}

\newcommand{\pp}{\mathcal G}

\newcommand{\IG}{\mathbb{I}}

\newcommand{\pto}{\boldsymbol{\pi}}  
\newcommand{\cpo}{\boldsymbol{\tau}}

\newcommand{\dom}{ \unlhd}

\newcommand{\bps}{\C_b} %\C^{(2)}}

\newcommand{\pps}{\C_p}%^{(>2)}}

\newcommand{\lift}{L}

\newcommand{\rank}{\rho}

\newcommand{\glb}{\mathop{glb}}

\usepackage{xcolor}  
\title[Encoding and ordering $X$-cactuses]{Encoding and ordering $X$-cactuses}

\author{Andrew Francis}
\address{Centre for Research in Mathematics and Data Science, Western Sydney University, Sydney, Australia\\
	E-mail address: a.francis@westernsydney.edu.au}
\author{Katharina T. Huber, Vincent Moulton \and Taoyang Wu}
\address{School of Computing Sciences, University of East Anglia,\\ Norwich, NR4 7TJ, UK\\ E-mail addresses: \{k.huber,v.moulton,taoyang.wu\}@uea.ac.uk}

\begin{document}
\begin{abstract}
Phylogenetic networks are a generalization of evolutionary or phylogenetic trees
that are commonly used to represent the evolution of species which 
cross with one another. A special type of phylogenetic 
network is an {\em $X$-cactus}, which is essentially a cactus graph 
in which all vertices with degree less than three are labelled
by at least one element from a set $X$ of species.
In this paper, we present a way to {\em encode} $X$-cactuses in terms
of certain collections of partitions of $X$ that naturally arise from
$X$-cactuses. Using this encoding, we also introduce a partial
order on the set of $X$-cactuses (up to isomorphism), and 
derive some structural properties of the resulting partially ordered set.
This includes an analysis of some properties of
its least upper and greatest lower bounds.
Our results not only extend some fundamental properties of  
phylogenetic trees to $X$-cactuses, but also provides 
a new approach to solving topical problems in 
phylogenetic network theory such as deriving consensus networks.\\

\noindent{\bf keyword} X-cactus, poset, bound, consensus network, supernetwork, phylogenetic network
%\MSC 05C05 \sep 05C20 \sep 05C85 \sep 05D15 \sep 92D15
%\end{keyword}
\end{abstract}

\maketitle

\section{Introduction}

In this paper, we let $X$ denote a finite, non-empty set. 
An {\em $X$-tree} $\TT=(T,\phi)$ is a graph theoretical tree $T=(V,E)$ together with a 
map  $\phi:X \to V$ whose image includes all vertices in $T$ with degree 
two or less. In case $\phi$ is a bijection onto the leaf-set of $T$, $T$
is called a {\em phylogenetic tree}.  $X$-trees naturally arise in evolutionary biology where 
they are commonly used to represent the
evolution of a set $X$ of species \cite{semple2003phylogenetics}. 
A fundamental property of $X$-trees is that a partial 
order $\leq$ can be defined on the set $\mathcal T(X)$ of $X$-trees (up to isomorphism)
by defining $\TT \le \TT'$ for two trees $\TT, \TT'\in\TT(X)$ precisely if a subset of edges in $\TT'$ can 
be contracted so as to obtain $\TT$ \cite[Section 3.2]{semple2003phylogenetics}. 
The poset $(\mathcal T(X), \le)$  has several interesting 
structural properties, some of which have proven useful in 
developing new insights and methodologies in phylogenetics. For example, 
lower bounds in $(\mathcal T(X), \le)$
correspond to consensus trees \cite[Section 3.6]{semple2003phylogenetics},
which are used in phylogenetics to summarise large collections
of phylogenetic trees \cite{bryant2003classification}.
	
\begin{figure}[ht]
	\begin{center}
	\includegraphics[width=0.7\textwidth]{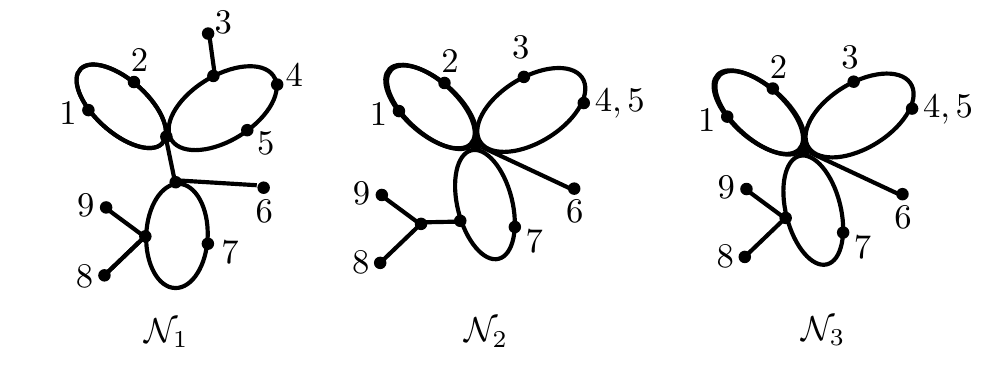}
	\caption{Three $X$-cactuses $\N_1, \N_2, \N_3$ on the set $X=\{1,\dots,9\}$.
		$\N_3\leq\N_1$ as $\N_3$ can be obtained from $\N_1$ 
		by contracting two cut edges
		and an edge in the top-right cycle. Also, $\N_3\leq \N_2$ as  
		$\N_3$ can be obtained from $\N_2$
		by  contracting a cut edge. 
		%In fact $\mathcal N_3$ is a greatest lower bound
		%for $\{\N_1,\N_2\}$ in $\pp(X)$ for $\leq$.
	}
	\label{f:1_Fig_poset}
	\end{center}
\end{figure}

Recently, there has been a great deal of interest in phylogenetic networks,
a generalization of phylogenetic trees
that are used to represent the evolution of species which cross with one another,
such as plants and viruses \cite{huson2010phylogenetic}.  An important 
class of such networks is the collection of $X$-cactuses \cite{hayamizu2020recognizing} (also known as 
1-nested networks \cite{gambette2017uprooted}), which
contains the well-known subclass of {\em level-1 networks} \cite{gambette2012encodings}.  
A {\em cactus} is a connected graph $N$
such that any two distinct cycles in $N$ share at most one vertex;
an {\em $X$-cactus} $\N=(N,\phi)$
is a cactus $N=(V,E)$ together with a map  $\phi:X \to V$ whose 
image includes all vertices in $N$ with degree two or less
(e.g., see Figure~\ref{f:1_Fig_poset}).
Note that an $X$-tree is simply an $X$-cactus whose underlying graph is a tree.
In this paper, we show that by considering edge-contractions for $X$-cactuses we
can obtain a partial order $\le$ on the set $\pp(X)$
of $X$-cactuses (up to isomorphism) 
that naturally extends the edge-contraction ordering on $X$-trees. 
As well as studying structural  properties of the poset $(\pp(X),\le)$ 
we show that, as with $X$-trees, we 
can use the ordering to define consensus networks for $X$-cactuses, a problem 
that is of topical interest in the theory of phylogenetic networks. 

We now describe the contents of the rest of this paper, at
the same time giving an overview of our main results. After presenting some 
preliminaries in Section~\ref{sec:preliminaries}, in Section~\ref{sec:encoding} we 
describe a way to {\em encode} $X$-cactuses. 
To help put this statement into context, we first recall that a
fundamental property of an $X$-tree $\TT$ is that it is completely
determined by the set $\Sigma(\TT)$ of bipartitions on $X$ that
is obtained by removing precisely one edge from $\TT$ for each edge in $\TT$.
More precisely, the {\em Splits-Equivalence Theorem for $X$-trees} states that 
if $\Sigma$ is a set of bipartitions of $X$, then 
there is an $X$-tree $\TT$ such that  $\Sigma=\Sigma(\TT)$ 
if and only if $\Sigma$ satisfies a certain pairwise condition 
called compatibility, in which case $\TT$ is the unique such $X$-tree
up to isomorphism \cite{buneman1971recovery}
(see also \cite[Thoerem 3.1.4]{semple2003phylogenetics}). 

To obtain our encoding for $X$-cactuses,  we consider the removal 
of either a cut edge or of all edges in some cycle in an $X$-cactus. The
first removal results in a bipartition of $X$, just as with $X$-trees (e.g. 
removal of the interior cut edge in the cactus $\mathcal N_1$ in Figure~\ref{f:1_Fig_poset}
gives the bipartition $\{\{1,2,3,4,5\},\{6,7,8,9\}\}$).
The second removal results in a partition of $X$ whose size is
the length of the cycle -- clearly, the ordering
of the vertices in the cycle is also important, and so we define the 
concept of a {\em circular partition} or a partition with a 
circular ordering, to capture this fact (e.g. 
removal of the edges in the top right cycle in the 
cactus $\mathcal N_2$ in Figure~\ref{f:1_Fig_poset}
gives the circular partition $(\{1,2,6,7,8,9\},\{3\},\{4,5\})$).
In Theorem~\ref{t:bijection.QX.CN}, we show
that an $X$-cactus $\N$ is encoded by its corresponding set $\mathcal C(\N)$
of circular partitions, and characterise when an arbitrary collection of 
circular partitions corresponds to
a (necessarily unique) $X$-cactus. As with $X$-trees, 
this characterization is  given in terms of a pairwise condition 
arising from the notion of {\em strongly compatibility}, a 
concept that was introduced in \cite{dress1997trees}.

In Section~\ref{sec:poset}, we define an order $\le$ on the 
set $\pp(X)$ of $X$-cactuses $\pp(X)$ (up to isomorphism). As
with $X$-trees, this is essentially defined via edge contraction, 
where one network is less than another if it can 
be obtained by contracting a subset of edges in the first (see e.g.  Figure~\ref{f:1_Fig_poset}). 
Some care is needed however in case an edge in a 3-cycle is contracted; 
we define a so-called {\em triple contraction} to deal with this situation.
In Theorem~\ref{l:partial.order.on.QX}, we show that $(\pp(X),\le)$ is a poset, 
and present some of its structural properties. 
In particular, we show that $\pp(X)$ is a graded poset with a unique minimal element (the 
$X$-cactus whose underlying graph is a single vertex), 
and characterize its maximal elements (Theorem~\ref{l:partial.order.on.QX}). 
In Section~\ref{sec:characterize}, we show that
the poset $(\pp(X),\le)$ can also be given in terms
of  $X$-cactus encodings. In particular, we show that $\N \le \N'$ 
holds for any two $\N,\N'\in\pp(X)$ if and
only if $\C(\N)$ can be mapped in a special 
way into $\C(\N')$ (Theorem~\ref{thm:poset:net:partition}).

In Sections~\ref{sec:upper} and \ref{sec:lower},
we consider upper and lower bounds in the poset $(\pp(X),\le)$. 
In general, these bounds have a more
complicated behaviour than upper and lower bounds in the poset $(\mathcal T(X),\le)$.
Indeed, unlike the poset of $X$-trees, there may exist non-unique 
least upper and greatest lower bounds. Even so, we are
able to characterize upper and lower bounds in  $(\pp(X),\le)$ (Corollary~\ref{cor:bound:char}).
In addition, we shed some light on least upper bounds
for pairs of $X$-cactuses (Theorem~\ref{thm:up}), and characterize
greatest lower bounds for arbitrary sets of $X$-cactuses 
(Theorem~\ref{thm:glb}).  We expect that this latter result could be a useful 
starting point for developing methods to find consensus networks
for collections of $X$-cactuses. 
In Section~\ref{sec:discussion}, we conclude by presenting some 
open problems and new directions.

\section{Preliminaries}\label{sec:preliminaries}

\subsection{Graphs and $X$-cactuses}
All graphs in this paper are {\em undirected} and {\em simple}, that is, they contain neither loops 
nor parallel edges. Let $G$ be a graph. A {\em leaf} in $G$ is a vertex with
degree one and an {\em internal vertex} of $G$ is a vertex that is not a leaf.
A {\em path} in a graph $G$ is a sequence of distinct vertices $v_1,v_2,\cdots,v_m$ 
such that $v_i$ is adjacent to $v_{i+1}$ for $1 \le i < m-1$. If, in addition, $v_1$ and $v_m$ 
are adjacent, then the subgraph of $G$ whose vertex set is $\{v_1,\cdots,v_m\}$ and whose 
edge set consists of $\{v_1,v_m\}$ and $\{v_i,v_{i+1}\}$ for $1\le i <m-1$ is a {\em cycle}.
% (taking  indices$\mod m$).  
A cycle is called {\em tiny} if it contains precisely 
three vertices. A {\em block} of $G$ is a maximal subgraph of $G$ not containing a
cut vertex, and the set of blocks of $G$ is denoted by $\B(G)$. 
A graph is {\em trivial} if it contains only one vertex, and {\em nontrivial} otherwise. 
Note that the trivial graph is defined as having no blocks.

A {\em cactus} is a connected graph  such that any 
two distinct cycles in it share at most one vertex. 
Equivalently, a cactus $N$ is a connected graph in which each edge 
belongs to at most one cycle so that, in particular, $N$ is a cactus if and only if
each edge in $N$  belongs to one and only block in $\B(N)$. Note that 
each block in a cactus is either a cut edge or a cycle, and that
the trivial graph is the only cactus that does not contain any block.

Now, for $X$ a non-empty finite set, an {\em $X$-cactus}, is 
a pair $\N=(N,\phi)$ where $N=(V,E)$ is a cactus and  $\phi:X\to V$
is a {\em labelling map}, i.e. a 
map from $X$ to $V$ such that every  vertex of degree at most two in $N$ is contained in its image. 
To help reduce notational complexity, in case there is little chance for confusion 
we shall just extend graph theoretical concepts and notations to $X$-cactuses in the natural way.
As mentioned in the introduction, an $X$-tree is an $X$-cactus whose underlying graph is a tree.
%Note that the set of {\em $X$-trees}
%(as defined in e.g. \cite[Section 2.1]{semple2003phylogenetics}), 
%is a subset of a the set of $X$-cactuses -- it just consists of
%all $X$-cactuses $\N=(N,\phi)$ where $N$ is a tree.
Two $X$-cactuses $\N=((V,E),\phi)$ and $\N'=((V',E'),\phi')$ are {\em isomorphic} 
if there exists a bijective map $f: V \to V'$ such that (i) $\{u,v\}\in E$ if and only if 
$\{f(u),f(v)\}\in E'$ and (ii) for all $x\in X$ we have $f(\phi(x))=\phi'(x)$.
The set consisting of all $X$-cactuses up to isomorphism is denoted by $\pp(X)$.
An $X$-cactus $\N=(N,\phi)$ is called {\em trivial} if $N$ is trivial 
and it  is called {\em binary} if every internal vertex of $N$ has degree 3.
A {\em phylogenetic $X$-cactus} is an $X$-cactus in which the labelling map is a bijection
onto its leaves.

We shall use two basic graph operations in this paper 
which are defined as follows. 
Given a graph $G=(V,E)$ and a subset $E' \subseteq E$, we let $G - E'$ 
be the graph with vertex set $V$ and edge set $E- E'$.
In case $E'=\{e\}$ we denote $G - E'$ by $G - e$.
In addition, if $v \in V$, we  let $G- v$ be the graph 
obtained from $G$ by deleting $v$ and all the edges incident 
with $v$. Finally, for a vertex $v$ in $G$ with degree two, 
the graph $G'=(V',E')$ obtained from $G$ by {\em suppressing} $v$ has 
vertex set $V'=V-\{v\}$ and edge set $E'=(E-\{\{u_1,v\},\{u_2,v\}\}) \cup \{u_1,u_2\}$, where 
$u_1 \neq u_2$ are adjacent to $v$ in $G$. Note that suppressing 
a degree two vertex $v$ in  a cactus decreases the number 
of its edges either by one or by two; the latter occurs 
if and only if $v$ is contained in a tiny cycle. 
% (when $v$ is contained in a tiny cycle). 

\subsection{Circular orderings}  

Given a finite set $Y$ with $m\ge 2$ elements,  and a linear order 
$\theta=(y_1,y_2,\cdots,y_m)$ of $Y$, we let $s(\theta)=(y_2,y_3,\cdots,y_m,y_1)$
and $r(\theta)=(y_m,y_{m-1},\cdots,y_2,y_1)$. 
Two linear orderings $\theta$ and $\theta'$ are {\em (circular) equivalent}, 
denoted by $\theta\sim \theta'$, if either $\theta=\theta'$ or there exist $k+1$ 
linear orderings $\theta_0=\theta$, $\theta_k=\theta'$ ($k \ge 1$) such 
that for $0 \le i <k$, we have $\theta_{i+1}=r(\theta_i)$ or $\theta_{i+1}=s(\theta_i)$.
Using the fact that $s^m(\theta)=\theta=r^2(\theta)$, it is straightforward to check that $\sim$ 
is an equivalence relation on the set of linear orderings of $Y$. 

A \emph{circular ordering} of $Y$
is an equivalence class of $\sim$. In particular, if 
$\theta$ is a linear ordering of $Y$, then $[\theta]$ is 
the equivalence class consisting of  
the distinct linear orderings of $Y$ that are equivalent to $\theta$. 
For example, for $Y=\{1,2,3,4\}$,
 \begin{align*}
[(1,3,2,4)]=\{ &(1,3,2,4),(3,2,4,1),(2,4,1,3),(4,1,3,2),\\
&(1,4,2,3),(4,2,3,1),(2,3,1,4),(3,1,4,2)\}.
\end{align*}
Intuitively, in case $m \ge 3$, a circular ordering of $Y$ is a 
labelling of the vertices of a regular $m$-gon by the elements in $Y$.
%Two elements $a, b \in Y$  are adjacent in a circular ordering $[\theta]$ if there 
%exists a linear ordering of form $(a,b,\dots)$ that is equivalent to $\theta$. 
 Indeed, the operations $s$ and $r$ can be seen as the generators of the
 dihedral group $D_m$ acting on an $m$-gon ($s$ a rotation and $r$ a
 reflection), and the equivalence classes of $\sim$ describe the orbits of
 this action.

\subsection{Circular partitions} 

Recall that a {\em partition} $\pi$ of set $X$ with $|X|\geq 2$ 
is a set consisting of at least two
nonempty pairwise disjoint proper subsets of $X$ whose union is equal to $X$. 
Each element in $\pi$ is  called a {\em part} of $\pi$ and 
the {\em size} of $\pi$ is its number of parts.
For example, $\pi=\{\{1,3\},\{2\},\{5\},\{4\}\}$ is a partition of $\{1,2,3,4,5\}$ with four parts; 
we shall also denote partitions such as $\pi$ by $13|2|5|4$, where the order of 
listing the parts does not matter. 
We refer to a partition as a {\em split} if it is of size two 
and a {\em proper partition} otherwise.
The set of partitions of $X$ is denoted by $\Pi(X)$.
Two partitions $\pi$ and $\pi'$ in $\Pi(X)$
are \emph{compatible} (also known as strongly compatible \cite{dress2010topology}) if
there exists a part $A\in\pi$ and a part $B\in\pi'$ such that $A\cup B=X$, and 
{\em incompatible} otherwise. 
Note that it follows that 
a split is compatible with itself while a proper partition is incompatible with itself.

A \emph{circular partition} of $X$ is an ordered pair $\sigma=(\pi,\tau)$ 
where  $\pi$ is a partition of $X$
and $\tau$ a circular ordering of the parts in $\pi$. 
We often use $[A_1|A_2|\cdots|A_k]$ to denote a circular partition consisting 
of the partition $\pi=A_1|A_2|\cdots |A_k$, $k \ge 2$, of $X$ and 
the circular ordering $\tau=[(A_1,A_2,\cdots,A_k)]$. 
Note that if $|\pi|=2$, i.e. $\pi$ is a split, then there exists only one possible circular 
ordering of $\pi$, and so we shall just call such a circular partition a split. We call a
circular partition $[A_1|A_2|\cdots|A_k]$ {\em proper} if $k\ge 3$,  and we call a collection 
of circular partitions {\em proper} if every partition in the collection is proper. 
Given a circular partition $\sigma=(\pi,\tau)$ of $X$ we let $\underline{\sigma} =\pi$ 
and call $\underline{\sigma}$ the partition {\em induced} by  $\sigma$. 
For example, $\sigma_1=[13|2|5|4]$ and $\sigma_2=[2|13|5|4]$ are two 
distinct circular partitions of $\{1,2,3,4,5\}$ because
$[(\{1,3\},\{2\},\{5\},\{4\})]\not= [(\{2\},\{1,3\},\{5\},\{4\})]$. Note however that $\underline{\sigma_1}=\underline{\sigma_2}=13|2|4|5$. For a set $\C$
of circular partitions, let $\Pi(\C):=\{\underline{\sigma}\,:\,\sigma\in \C\}$ be the set 
of partitions induced by $\C$. We shall also let $\C(X)$ denote
the set of circular partitions of $X$, $\bps(X)$
the set of splits of $X$, and $\pps(X)$ the set of 
proper circular partitions of $X$.  
%For example, the  circular
%partitions $\sigma_1=[13|2|5|4]$ and $\sigma_2=[2|13|5|4]$
%are distinct because their linear orderings, 
%and they are both contained in $\C_p(\{1,\ldots,5\})$.

Given a proper circular partition of $X$, we can {\em merge} any two adjacent parts into 
one part to construct another circular partition of $X$. For example, merging 
the two adjacent parts $\{2\}$ and $\{5\}$ in 
$\sigma_1=[13|2|5|4]$  results in the circular partition $[13|25|4]$. 
Given two circular partitions $\sigma$ and $\sigma'$ of $X$
we set $\sigma' \preceq \sigma$, if either 
(i) $\sigma=\sigma'$ or (ii)  $\sigma'$ is proper and $\sigma'$ 
can be obtained from $\sigma$ by applying a (necessarily finite) 
sequence of merges. 
%Note that the concept of refinement is introduced in a way to be consistent 
%with properness, that is, a split is only refined by itself, 
%while a proper circular partition is only refined by a 
%proper circular partition.
Furthermore, let $\sigma'\prec \sigma$ denote $\sigma'\preceq \sigma$ and $\sigma\not =\sigma'$. 
Note that if $\sigma'$ is a split and 
$\sigma' \preceq \sigma$ or $\sigma\preceq \sigma'$, then  $\sigma = \sigma'$. Also, if $\sigma'$ is not a split, $|\underline{\sigma}|=3$ and $\sigma'\preceq \sigma$ then $\sigma=\sigma'$. 
Moreover,  $\sigma\prec \sigma'$ implies that $\sigma$ and $\sigma'$ are incompatible. 
It is straight-forward to check that $\preceq$ is 
reflexive, antisymmetric, and transitive, and so $(\C(X),\preceq)$ is a poset. 
%Here we will use the convention that the bottom element in this poset is the trivial partition. 
%the circular partition with one part which is denoted by $[X]$ and referred 
% to as the {\em degenerate} partition.

Two circular partitions $\sigma$ and $\sigma'$ of $X$ are {\em compatible} if $\underline{\sigma}$ and $\underline{\sigma'}$ are compatible, and {\em incompatible} 
otherwise. 
Note that this definition implies that a circular partition  $\sigma$ on $X$ is compatible with itself if and only if $\sigma\in \C_b(X)$. 
A set $\C$ of circular partitions is compatible if each pair 
of distinct circular partitions in $\C$ is compatible. Here we use the convention that the emptyset of 
circular partitions is compatible.
Note that the circular orderings of two compatible proper partitions 
are `consistent' in the following sense.

\begin{lem}
	\label{lem:ord:compatible}
	Suppose that two proper circular partitions $\sigma_1$ and $\sigma_2$ of $X$ are  compatible. 
	Then there  exists a circular partition $\sigma'$ in $\C(X)$ with $\sigma_1 \preceq \sigma'$ 
	and $\sigma_2 \preceq \sigma'$.
\end{lem}

\begin{proof}
Since $\sigma_1$ and $\sigma_2$ are compatible, there exists a part $A_1$ in $\underline{\sigma_1}$ 
and $B_1$ in $\underline{\sigma_2}$ such that $A_1\cup B_1=X$. Without 
loss of generality, we may assume that $\sigma_1=\big[A_1|A_2|\dots|A_t \big]$ 
and $\sigma_2=\big[B_k|B_{k-1}|\dots |B_1 \big]$ for some $t,k\ge 3$. Since $A_1\cup B_1=X$, 
it follows that $B_i\subseteq A_1$ for $2\le i \le k$ and $A_i\subseteq B_1$ for $2\le j \le t$.

Now assume first that $A_1\cap B_1\not =\emptyset$. Then  
$\sigma'=\big[ B_k|B_{k-1}|\dots B_2 |B_1\cap A_1| A_2|\dots|A_t \big]$
is a proper circular partition of $X$, and it is straightforward to check that 
$\sigma_1 \preceq \sigma'$ and $\sigma_2  \preceq \sigma'$. 
In case $A_1\cap B_1 =\emptyset$, the same relationships
hold for $\sigma'=\big[ B_k|B_{k-1}|\dots |B_2| A_2|\dots|A_t \big]$.	
\end{proof}

\subsection{Tree representations of partitions}

\begin{figure}[ht]
	\begin{center}
	\includegraphics[width=0.9\textwidth]{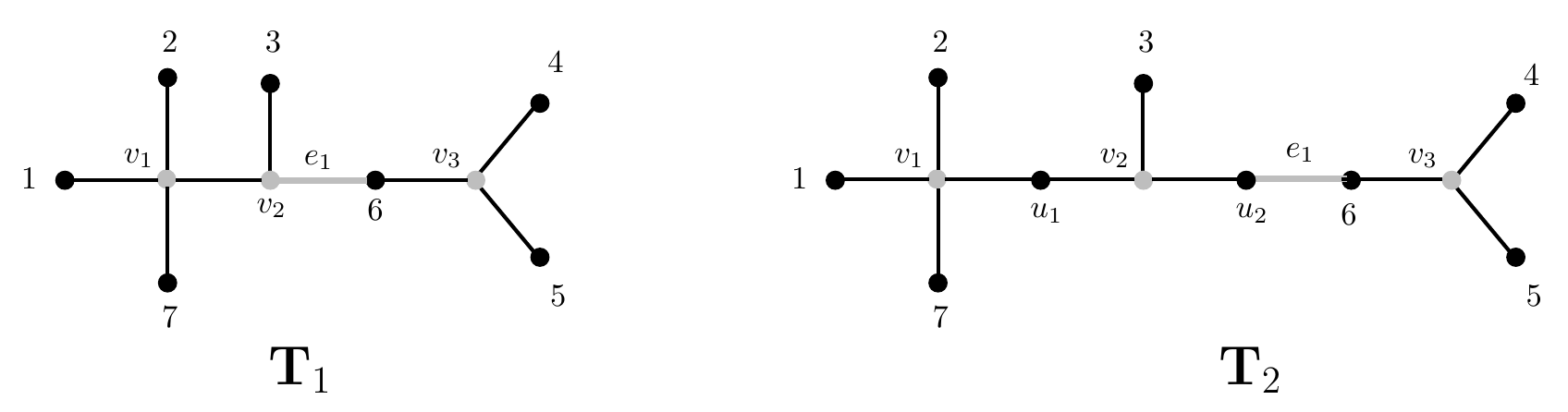}	
	\caption{Two semi-labelled $X$-trees $\T_1$ and $\T_2$.
		 For the set  $\Pi=\{\pi_1,\pi_2,\pi_3,\pi_4\}$ 
		 %of compatible partitions on $X=\{1,\dots,7\}$ 
		 with  $\pi_1=1|2|7|3456$, $\pi_2=127|3|456$, $\pi_3=12367|5|4$, $\pi_4=1237|645$ 
		and $\kappa_j(\pi_i)=v_i$ for $1\le i \le 3$ and $j=1,2$ and  $\kappa_j(\pi_4)=e_1$ for $j=1,2$,
		the pair $(\mathbf{T}_1,\kappa_1)$ is a semi-tree representation of $\Pi$ and the pair $(\T_2,\kappa_2)$ is the perfect semi-tree representation
		of $\Pi$. In each case $Im(\kappa)$
		is indicated in grey. Furthermore, $h(e_1,\kappa_1)=2$ and $h(e_1,\kappa_2)=1$.
	} 
	\label{f:tree:rep}
	\end{center}
\end{figure}

For later use, we recall some definitions and results on tree representations of partitions 
developed in~\cite{dress2010topology}.  To this end, we need to generalize
the concepts for $X$-cactuses to semi-labelled $X$-cactuses.
Formally, a \emph{semi-labelled $X$-cactus $\bN=(N,\psi)$}  is an ordered 
pair where $N=(V,E)$ is a cactus and $\psi:X\to V$ is a map such 
that every leaf in $N$ is  contained
in $\psi(X)$. Note that a semi-labelled cactus $\bN=(N,\psi)$ 
is an $X$-cactus if and only if every degree two 
vertex in $N$ is contained in $\psi(X)$.  Note that if $N$ is a 
tree, the pair $(N,\psi)$ is also referred to as a {\em semi-labelled $X$-tree}.
See Figure~\ref{f:tree:rep} for two examples of semi-labelled $X$-trees.

Given an unlabelled cut vertex $v \in V$ in a semi-labelled
$X$-cactus $\bN=(N=(V,E),\phi)$ (i.e. some cut vertex 
$v$ of $N$  not contained in $ \phi(X)$), let  $\pto(v)$ be the partition of $X$ induced by $\bN  -v$. 
In other words, $A$ is a part in $\pto(v)$ if and only if for all pairs of (not necessarily distinct) 
vertices $x,x' \in A$, no path between $x$ and $x'$ in $N$ 
contains $v$. Similarly, for any cut edge $e$ of $N$, let $\pto(e)$ 
be the split of $X$ induced by $N - e$. For example,
For example, in Figure~\ref{f:tree:rep} the 
partition induced by $\mathbf{T}_1 - v_2$ is given by $127|3|456$.

We define a {\em semi-tree representation} of a collection $\Pi$ 
of partitions of  $X$ to be an  ordered pair $(\T, \kappa)$ consisting of a semi-labelled $X$-tree $\T=(T=(V,E),\psi)$  
and a (necessarily injective) map $\kappa: \Pi \to (V-\psi(X))\cup E$ 
such that for each partition $\pi \in \Pi$, the image $\kappa(\pi)$ 
is either an unlabelled vertex of degree at least three or an edge, 
and $\pi=\pto(\kappa(\pi))$ holds. 
If $\Pi=\emptyset$, then
we use the convention that $\kappa$ is the {\em empty function}, that
is, the image $Im(\kappa)$ of $\kappa$ is the empty set.
Note that if $\Pi\not=\emptyset$ then $T$ must contain at least one edge. For each edge $e=\{u,v\}\in E$,
let $h(e,\kappa)$ be the number of partitions $\pi$ in $\Pi$ with $\kappa(\pi) \in \{u,v,e\}$. 
Note that $0\le h(e,\kappa) \le 3$. 
The representation  $(\T, \kappa)$ is \emph{perfect} if $h(e,\kappa)=1$ for all 
edges $e$ in $\T$ 
and $\kappa(\pi)$ is an edge in $\T$ if and only if $\pi$ is a 
split (see e.g. Figure~\ref{f:tree:rep}). We use the convention  that
a semi-tree representation whose underlying $X$-tree is trivial is perfect.
%  In case $\T$ is an $X$-tree, 
% $(\T,\kappa)$ is referred to as an $X$-tree 
% representation of $\Pi$ (see Figure~\ref{f:tree:rep}(i) for an example).

We now prove a simple but useful 
extension of \cite[Theorem 4]{dress2010topology} concerning
semi-tree representations.

\begin{thm}
\label{semi-tree:representation}
Suppose that $\Pi$ is a collection of partitions of $X$. Then $\Pi$ is compatible if and only if
there exists a semi-tree representation $(\T, \kappa)$ of $\Pi$. Moreover, if $\Pi$ is
compatible, then there exists a (necessarily unique) perfect semi-tree 
representation $(\T, \kappa)$ of $\Pi$.
\end{thm}

\begin{proof}
Clearly if there exists a semi-tree representation 
of $\Pi$, then $\Pi$ is compatible. 

Conversely, suppose $\Pi$ is compatible.
If $\Pi=\emptyset$, then  the theorem holds 
since $(\bf T,\kappa)$ is a semi-tree representation for $\Pi$, where
$\bf T$ is the semi-labelled trivial
$X$-tree and $\kappa$ the empty function on $\Pi$. 
So, assume $\Pi\not=\emptyset$.

In \cite[Theorem 4]{dress2010topology} it is proven
that, up to isomorphism, there exists a unique semi-tree representation $(\T,\kappa)$ of $\Pi$
for which $\T=((V,E),\phi)$ is an $X$-tree with $|E|\geq 1$ and $h(e,\kappa)>0$ holds for all every $e \in E$
(see e.g. Figure~\ref{f:tree:rep}). We now describe 
how to obtain a perfect semi-labelled 
representation $(\T'=((T',\phi'),\kappa')$ of $\Pi$  from $(\T,\kappa)$
which will complete the proof of the theorem. The
fact that $(\T'=((T',\phi'),\kappa')$ is perfect is straight-forward
to verify and so we omit this.

The tree $\T'$ is obtained from $\T$ by 
inserting $h(e,\kappa)-1$  additional unlabelled degree two 
vertices into each edge $e \in E$. 
The labelling map $\phi'$ is the same as $\phi$, i.e., 
for each $x\in X$, we let $\phi'(x)=\phi(x)$. 
The map $\kappa'$ is derived from $\kappa$ as follows: 
For $\pi\in \Pi$ with $|\pi|\ge 3$,   
$\kappa(\pi)$ is a vertex in $\T$ and we set  
$\kappa'(\pi)=\kappa(\pi)$. 
For $\pi\in \Pi$ with  $|\pi|=2$, 
$\kappa(\pi)$ is an edge $e=\{v_1,v_2\} \in E'$.
We define $\kappa'(\pi)$ 
depending on the value $h(e,\kappa)$. If $h(e,\kappa)=1$, 
then $e$ is also an edge in $E$, and we let $\kappa'(\pi)=\kappa(\pi)=e$. 
If $h(e,\kappa)=2$, then we subdivide $e$ 
into two edges $e_1$ and $e_2$, where the indices are chosen 
in such a way that $e_2$ is not incident with a 
vertex in $\kappa(\Pi)$ and we let $\kappa'(\pi)=e_2$.  
If $h(e,\kappa)=3$, then we subdivide $e$ into three edges $e_1$, $e_2$, and $e_3$, where 
the indices are chosen in such a way that $e_2$ is not incident with a 
vertex in $\kappa(\Pi)$ and we let $\kappa'(\pi)=e_2$. 
\end{proof}

\section{Encoding $X$-cactuses}\label{sec:encoding}

In this section, we introduce an encoding for $X$-cactuses
that is given in terms of circular partitions of $X$
We begin by describing a natural way to associate 
a collection of compatible 
circular partitions to an $X$-cactus $\N=(N,\phi)$.

Given an $X$-cactus $(N,\phi)$, 
define a vertex in $N$ to be {\em terminal} if it belongs to one and only 
one block of $N$.  Each terminal vertex in $N$ is of degree two or one, and hence 
contained in $\phi(X)$. Now suppose $N$ has at least two vertices,  and 
that $v$ is a vertex in a block $B$ of $N$.  Let $(N-B)_v$ be the 
connected component in $N -E(B)$ that contains $v$. Then  
$(N-B)_v$ contains at least one element in $\phi(X)$ because it contains at
least one terminal vertex. 

Now, to each block $B$ in $N$, we associate a circular partition $\cpo(B) \in \C(X)$
as follows. Let $v_1,\cdots,v_k$, $k\ge 2$, be a labelling of the vertices in $B$ so that $\{v_i,v_{i+1}\}$ 
is an edge in the block for each $1\le i \le k$, where the indices
are given modulo $k$. Let $V_i$ be the vertex set of the 
connected component $(N-B)_{v_i}$ for $1\le i \le k$. Then 
by the previous paragraph it follows that $V_i$ contains at least 
one element in $\phi(X)$. Furthermore, it is straightforward to 
check that $V_i\cap V_j=\emptyset$ for $1\le i< j\le k$, 
and $V(N)=\bigcup_{1\le i \le k}V_i$. Hence
$\cpo(B):=[\phi^{-1}(V_1)|\phi^{-1}(V_2)|\cdots |\phi^{-1}(V_k)]$ 
is a circular partition in $\C(X)$.  Note that if $k=2$, then $B$ is  
a cut edge and $\cpo(B)$ is a split. 
Furthermore, in general $\cpo(B)$ is determined by $B$, but not by the labelling
that we chose for its vertices, since any other labelling
of this form induces the same circular partition.

Let $\C(\N)=\{\cpo(B)\,:\, B\in \B(N)\}$. In case $\N$ is 
the trivial $X$-cactus (i.e. $N$ is a single vertex), we define $\C(\N)$ to be the emptyset. 
Clearly, $\C(\N)$ is the (necessarily disjoint) union 
of $\C_b(\N)=\C_b(X) \cap \C(\N)$ and $\C_p(\N)=\C_p(X) \cap \C(\N)$. In particular, $|\C(\N)|$ is the 
number of cycles in $N$ plus the number of cut edges.
We now show that the set $\C(\N)$ is compatible.
%For $\C$ a collection of circular partitions, let  
%$\Pi(\C)=\{\underline{\sigma}\,:\, \sigma\in \C\}$.

\begin{lem}\label{l:CN.compatible}
Suppose that $\N$ is a $X$-cactus. Then there 
is a perfect semi-tree representation of $\Pi(\C(\N))$.
In particular, $\C(\N)$ is a compatible set of circular partitions.
\end{lem}

\begin{proof}
Let $\Pi=\Pi(\C(\N))$
%, $\Pi_b=\Pi(\C_b(\N))$, $\Pi_p=\Pi(\C_p(\N))$, 
and put $\N=(N,\phi)$.  
If $\N$ is the trivial $X$-cactus, then
$C(\N)=\emptyset=\Pi(\C(\N))$. The lemma then follows since 
$\C(\N)$ is compatible and $(\bf T,\kappa)$ is a perfect semi-tree
representation for $\Pi(\C(\N))$, where $\bf T$ is the
semi-labelled trivial $X$-cactus and $\kappa$ is the empty function
on $\Pi(\C(\N))$. 

Now, assume $\N$ is not the trivial $X$-cactus.
We first construct a semi-tree representation $((T,\psi),\kappa)$ of $\Pi$
(see e.\,g.\,Figures~\ref{f:tree:rep} and \ref{f:net}).
Let $\B_p(N)$ be the set of blocks of $N$ that are 
cycles, and $\B_b(N)$ be the set of blocks of $N$ that are cut edges. 
Then $\B(N) = \B_b(N) \coprod \B_p(N)$, and 
$\C_b(\N)$ and $\C_p(\N)$ are precisely the set of circular partitions induced 
by the blocks in $\B_b(N)$ and $\B_p(N)$, respectively.  

\begin{figure}[ht]
	\begin{center}
	\includegraphics[width=0.6\textwidth]{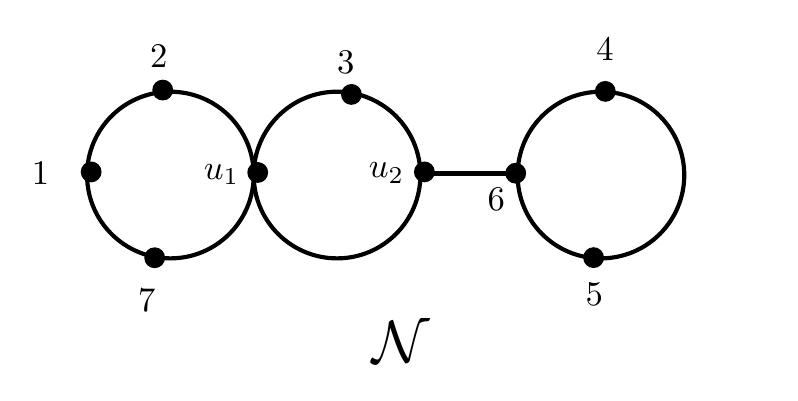}
	\caption{Example of an $X$-cactus $\N$ on $X=\{1,\dots,7\}$. The cactus 
		contains four blocks: cycles $B_1$, $B_2$ and $B_3$ (from left the right) 
		and the cut edge $B_4$. The set $\C(\N)$ contains four circular partitions 
		$\sigma_1=[1|2|3456|7]$, $\sigma_2=[127|3|456]$, $\sigma_3=[4|5|1236]$, 
		and $\sigma_4=[1237|456]$, where $\sigma_i=\cpo(B_i)$ for $1\le i \le 4$. Putting  $\sigma_i=(\tau_i,\pi_i)$ for all $1\leq i\leq 4$,
		 we have $\pi_i=\underline{\sigma_i}$ for all such $i$, and 
		the perfect semi-tree representation of $\Pi(\C(\N))=\{\underline{\sigma_i}\,:\, 1\le i \le 4\}$ 
		is depicted on the right of  Figure~\ref{f:tree:rep}.
		%,where $\pi_i=\underline{\sigma_i}$  for $1\le i \le 4$.
	} 
	\label{f:net}
	\end{center}
\end{figure}

We now construct $T$:
For every block $B$ in $\B_p(N)$, 
add a new vertex $v_B$, a new edge $\{v_B,v\}$ for 
each vertex $v$ in $B$,  
and remove all edges in $B$.  
Note that the vertex set $V(T)$ of $T$ is the 
disjoint union of the vertex set $V(N)$ of $N$ and the set $V^*(T)=\bigcup_{B\in \B_p(N)}\{v_B\}$ of new vertices. 
Furthermore, a vertex $v$ in $N$ is a leaf in $T$ if and only if $v$ is a 
terminal vertex in $N$. Finally, the set $E(T)\cap E(N)$ 
consists of all cut edges in $N$, one for each of the blocks in $\B_b(N)$.

Let $\psi$ be the labelling map from $X$ to $V(T)$ induced by 
$\phi$, that is, we have $v=\psi(x)$ for vertex $v$ in $T$ and  
$x\in X$ if and only if $v\in V(N)\subseteq V(T)$ and $v=\phi(x)$.  
Note that each leaf in $T$, as a terminal vertex in $N$, is necessarily 
contained in $\psi(X)$, and $V^*(T)\cap \psi(X)=\emptyset$. 

We now define the map $\kappa: \Pi \to \big(V(T)-\psi(X)\big)\cup E(T)$. 
If $\pi\in \Pi_p=\Pi(\C_p(\N))$, then there exists a unique circular partition $\sigma$ 
in $\C_p(\N)$ with $\pi=\underline{\sigma}$ and we let $\kappa(\pi)=v_B$, 
where $B$ is the unique block $B$ in $\B_p(N)$ with $\cpo(B)=\sigma$. 
%and $v_B$ is the vertex in $V^*(T)$ corresponding to $B$. 
Otherwise, $\pi \in \Pi_b=\Pi(\C_b(\N))$ and so there exists a unique 
split $\sigma$ in $\C_b(\N)$ with $\pi=\underline{\sigma}$. In that case,
we let $\kappa(\pi)=e_B$ where $B$ is the unique block $B$ in $\B_b(N)$ 
with $\cpo(B)=\sigma$ and $e_B$ is the unique edge in $B$ (which is 
a cut edge and hence contained in $E(T)\cap E(N)$). 

We show that $((T,\psi),\kappa)$ is perfect.
Since $\kappa(\pi)$ is an edge if and only if $\pi$ is a split, we only need to show that 
$h(e,\kappa)=1$ for each edge $e=\{u,v\}$ in $T$. 
To this end, it suffices to establish that there exists one and only one 
partition $\pi$ in $\Pi$ such that $\kappa(\pi)\in\{u,v,e\}$. 
This clearly holds if $e$ is contained in $E(T)\cap E(N)$ (i.e., $e$ is a cut edge in $N$).
Indeed, neither $u$ nor $v$ is contained in $V^*(T)= \kappa(\Pi_p)$ 
and $\kappa(\pi)=e$ holds if and only if  $\pi=\underline{\cpo(B_e)}$,
%, i.e. the partition corresponding to the circular partition $\cpo(B_e)$, 
where $B_e$ is the block consisting of the cut edge $e$. 
The other case is $e\in E(T)- E(N)$. Swapping $u$ and $v$ if necessary, we may 
assume that $u$ is contained in $V(T)\cap V(N)$, while $v$ is contained 
in $V^*(T)$. Hence $v=v_B$ for a (necessarily unique) block  $B$ in $\B_p(N)$. Then 
neither $u$ nor $e$ is contained in $\kappa(\Pi)$, and   $\kappa(\pi)=v$ if and 
only if $\pi=\underline{\cpo(B)}$.% is the partition corresponding to the circular partition $\cpo(B)$. 

The last statement of the lemma follows immediately from Theorem~\ref{semi-tree:representation}.
\end{proof}

\begin{figure}[ht]
	\begin{center}
	\includegraphics[width=0.4\textwidth]{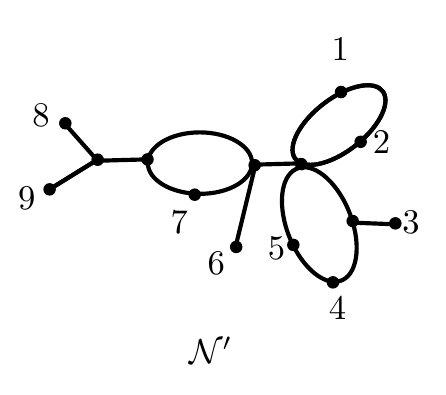}
	\caption{ The $X$-cactus $\N'=\N(\C)$ on $X=\{1,\ldots, 9\}$
		for the set $\C=\{\alpha,\beta, \gamma, s_1, s_2, s_3, s_6, s_8, s_9\}$ of circular partitions where $\alpha=[1|2|3456789]$,
		$\beta=[678912|3|4|5]$, $\gamma=[89|123456|7]$, $s_1=[12345|6789]$, $s_2=[89|1234567]$
		and $s_x = [x|X-x]$ for $x\in\{3,6,8,9\}$.
	} 
	\label{f:nc-construction}
	\end{center}
\end{figure}

We now describe a way to construct an $X$-cactus from a collection $\C \subseteq \C(X)$ of 
circular compatible partitions,  i.e., the reverse of Lemma~\ref{l:CN.compatible}. 
We first construct a labelled graph $\N(\C)=(N(\C),\phi)$ from $\C$. 
In case $\C=\emptyset$, we define $\N(\C)$ to be the trivial $X$-cactus. Otherwise,
for $\C \neq \emptyset$ we proceed as follows (see Figure~\ref{f:nc-construction}):
 
\begin{itemize}
\item[(i)] Let $(\T=(T,\phi),\kappa)$ be the perfect semi-tree representation 
of the set $\Pi=\Pi(\C)$ of partitions induced by $\C$
that is given by Theorem~\ref{semi-tree:representation}.
\item[(ii)] For every vertex $v$ in $\kappa(\Pi(\C_p))$, consider the proper circular 
partition $\sigma \in \C$ with $\pto(v)=\underline{\sigma}$. 
Give the vertices adjacent to $v$ a circular ordering 
$\lambda_v=[(v_1,v_2,\cdots,v_m)]$, $m \ge 3$, that
is consistent with the ordering of $\sigma$,
delete $v$, and add the edges $\{v_i,v_{i+1}\}$ for $1\le i \le m$, taking indices modulo $m$. 
Denote the resulting graph by $N=N(\C)$.
\item[(iii)] Define the labelling map $\phi: X \to V(N)$ to be the one 
naturally induced by the map $\phi: X \to V(T)$. 
\end{itemize}

Note that in case $\C$ consists solely of circular partitions that are splits, 
then, since each partition $\pi$ in $\Pi=\Pi(\C)$ is a split,
it follows that $\kappa(\pi)$ is necessarily an edge since Step (ii) does not apply, and 
hence no vertex in $N(\C)$ is contained in $\kappa(\Pi)$. 
In particular, $\N= \N(\C)$ is an $X$-tree,
and so $\N$ is an $X$-cactus with $\C(\N)=\C$.
We now show that $\N(\C)$ is an $X$-cactus for an arbitrary 
collection $\C$ of compatible circular partitions.

\begin{lem}
\label{lem:surjective}	
If $\C\subseteq \C(X)$ is a set of compatible circular partitions, then $\N(\C)$ is an $X$-cactus 
such that $\C(\N(\C))=\C$.
\end{lem}
\begin{proof}
The lemma holds in case $\C=\emptyset$ by convention, and if $\C$ contains only splits  then is holds
by the above remarks. So assume $|\C|>0$ and 
that $\C$ contains at least one proper circular partition.
Let $\Pi=\Pi(\C)=\{\underline{\sigma}\,:\,\sigma\in \C\}$ be 
the set of partitions on $X$ induced by $\C$.
Since $\C$ is compatible, it follows that $\Pi$ is also compatible. 

Consider the perfect semi-tree representation $(\T=(T,\phi),\kappa))$ of $\Pi$
given by Theorem~\ref{semi-tree:representation}. Then $E(T)|\geq 1$.
Let $V_o$ be the set of unlabelled degree two vertices in $T$. 
Let $E_1$ be the set of edges $e$ in $T$ 
for which $\kappa(\pi)=e$  holds for some split $\pi$  in $\Pi$.

Let $\C_p = \{\sigma_1,\dots,\sigma_k\}$, $k \ge 1$, be the subset of 
proper circular partitions in $\C$. Let 
$V_1=\{v_1=\kappa(\underline{\sigma_1}),\dots,v_k=\kappa(\underline{\sigma_k})\}$. 
Because $(\T,\kappa)$ is a perfect semi-tree representation, for each $v\in V_o$, 
there exists a vertex $v' \in V_1$ such that $v$ is adjacent to $v'$. Moreover,  
each edge $e$ in $E(T)-E_1$ is incident with a vertex $v$ in $V_1$. 

Let $\N_0=\T$. For $1\le i \le k$, let $\N_i$ be the graph obtained 
from $\N_{i-1}$ by performing Step (ii) in the construction of $\N(\C)$. For all $1\leq i\leq k$, 
let $N_i$ denote the underlying graph of $\N_i$. Note that, 
in particular, $N(\C) = N_k$. Since $v_i$ is a cut vertex in $N_{i-1}$ such that each 
connected component in $N_{i-1}-\{v_i\}$ contains 
precisely one vertex adjacent with $v_i$ in $N_{i-1}$, 
it follows that $N_{i}$ is a simple connected graph in which two cycles share at most one vertex. 
Moreover, $N_{i}$ contains precisely one more cycle, denoted by $C_i$, than $N_{i-1}$ 
and, by construction, $\cpo(C_i)=\sigma_i$.

Thus $N(\C) = N_k$ is a connected graph in which two cycles share at most one 
vertex with $\C_p(\N_k)=\C_p$.  Since each vertex $v$ in $V_o$ is adjacent to a 
vertex $v_i$ in $T$, by construction it follows that $v$ is a vertex in the 
cycle $C_i$ and hence $\N_k$ does not contain any unlabelled degree two vertices. 

Note that each edge added during the process of constructing $\N_k$ from $\N_0$ 
belongs to a cycle and each edge $e$ in $E(T) - E_1$ is removed as $e$ 
is incident with a vertex in $V_1$. Thus, the set of cut edges in $\N_k$ is $E_1$ and 
these cut edges induce precisely the splits that are contained in $\Pi$.  
It follows that $\N_k$ is an $X$-cactus 
with $\C(\N_k)=\C$, which completes the proof of the lemma.
\end{proof}

We now prove the main result of this section,  
an analogue of the Splits Equivalence Theorem  for $X$-trees
\cite[Theorem 3.1.4]{semple2003phylogenetics}. 

\begin{thm}\label{t:bijection.QX.CN}
Let $\C$ be a set of circular partitions of $X$.
Then there is an $X$-cactus $\N$ such that $\C = \C(\N)$ 
if and only if $\C$ is compatible. Moreover, if such an
$X$-cactus exists, then up to isomorphism, it is unique.
\end{thm}
\begin{proof}
	We may assume that $\C\not=\emptyset$ since otherwise the theorem holds.
We claim that the map which takes any $X$-cactus $\N \in \pp(X)$ to
the set $\C(\N)$ of circular partitions  induces 
a bijection between $\pp(X)$ and the 
set of compatible sets of circular partitions of $X$.
The theorem then follows immediately from this claim.

By Lemma~\ref{lem:surjective}, the map in the claim 
is surjective, and so it suffices to show that it is injective.
To this end, suppose that $\N$ and $\N'$ are two $X$-cactuses 
with $\C(\N)= \C(\N')$. We show that $\N$ 
is isomorphic to $\N'$, which will complete the proof of the claim.

By Lemma~\ref{l:CN.compatible}, there 
is a unique perfect semi-tree representation  $(\T,\kappa)$ of $\Pi(\C(N))$. 
Moreover, if we apply Steps~(i)--(iii) to $(\T,\kappa)$ to construct the network $\N(\C(\N))$
it is straightforward to check by considering the 
construction of $(\T,\kappa)$  used in the proof of  Lemma~\ref{l:CN.compatible} 
that the network $\N(\C(\N))$ that we obtain is isomorphic to $\N$.
But the same argument holds for $\N'$, and so it follows that 
$\N$ is isomorphic to $\N'$.
\end{proof}

\begin{rem}
	By Theorem~\ref{t:bijection.QX.CN}, the function 
	$d:\pp(X)\times \pp(X)\to \mathbb R_{\geq 0}$ given by
	$d(\N,\N')= |\C(\N) \Delta \C(\N')|$ for all $\N, \N' \in \pp(X)$ is
	a metric on the set $\pp(X)$. This can be regarded as a generalization
	of the well-known Robinson-Foulds metric (cf. \cite [p.25]{steel2016phylogeny}). 
\end{rem}

\section{A partial order on $X$-cactuses}\label{sec:poset}

In this section, we introduce a partial order $\le$ on the set $\pp(X)$ 
of  $X$-cactuses, and describe some of its basic properties. 
%The ordering extends the definition of the well-known partial order on $X$-trees
%mentioned in the introduction (c.f. \cite[Section 3.2]{semple2003phylogenetics}). 
We begin by defining the following two operations on an $X$-cactus
which are related to local subnetwork operations given in \cite{huber2016spaces}. 

Suppose that $\N=(N,\phi)$ is a non-trivial 
$X$-cactus. First, given any edge $e=\{u,v\} \in E(N)$
that is not contained in a tiny cycle,  an \emph{edge 
contraction (of $\N$ on $e$)} results in an $X$-cactus $\N'$ obtained by deleting $e$, 
identifying $u$ and $v$ as a new vertex and labelling that vertex by  
the elements in $\phi^{-1}(u)\cup \phi^{-1}(v)$.  Second, given
a tiny cycle $C=v_1,v_2,v_3$ in $\N$, a \emph{triple contraction 
	(of $\N$ on $C$)}  
results in an $X$-cactus $\N'$ obtained from $\N$ by deleting all  
vertices and edges in $C$, adding a new vertex $v'$ which we
label $\bigcup_{i=1}^3\phi^{-1}(v_i)$ and replacing each
edge $\{v_i,u\}$, $u\in V(N)-\{v_1,v_2,v_3\}$ by a new edge $\{u,v'\}$.

We now introduce the partial order $\le$ on $\pp(X)$.
We define a {\em contraction} on a non-trivial 
$X$-cactus in $\pp(X)$ to be either an edge
contraction or a triple contraction. For two $X$-cactuses
$\N ,\N'\in\pp(X)$ we then 
put $\N\le \N'$ if and only if there is a (possibly empty) sequence of 
contractions transforming $\N'$ to $\N$. 
Furthermore, we put $\N<\N'$ if $\N\leq \N'$ and $\N\not=\N'$.

We now present some basic properties of the ordering $\le$ on $\pp(X)$.
For the reader's convenience, we recall two key concepts from
poset theory (see e.g.~\cite[p.4--5]{birkhoff1940lattice}). 
Suppose $(S,\le)$ is an arbitrary poset. Then $s' \in S$ is a {\em coatom} of $s \in S$ (or $s$ {\em covers} $s'$)
if $s' < s$ and there is no element $s'' \in S$ such that $s' < s'' < s$.
The poset $(S,\le)$ is {\em graded} if it has a {\em rank function}, that 
is a map $\rho$ from $S$ into the integers 
which, for $s, s' \in S$,  satisfies (a) if $s' < s$, then $\rho(s') < \rho(s)$, 
and (b) if $s'$ is a coatom of $s$,  then $\rho(s) = \rho(s') +1$.

\begin{thm}\label{l:partial.order.on.QX}
Assume $|X|\ge 2$. Then the following statements hold:
\begin{itemize}
	\item[(i)] $\le$ is a partial order on $\pp(X)$.
	\label{p:part:poset}
	
	\item[(ii)]For all $\N_1,\N_2\in \pp(X)$,  $\N_1$ is a coatom of $\N_2$ if and only if $\N_1$ can be obtained from $\N_2$ by one contraction. 
	\label{p:part:coatom}
	
 	\item[(iii)] The order $\le$ has a  unique minimal element, namely 
 	the trivial $X$-cactus.
	\label{p:part:poset.unique.min.elt}

	\item[(iv)] The maximal elements under $\leq$ are the binary phylogenetic $X$-cactuses in which 
	every internal vertex is contained in some cycle.
		\label{p:part:poset.max.elts}

%An $X$-cactus $\N\in\pp(X)$ is {\em saturated} if it is a binary phylogenetic cactus in which 
%each non-leaf vertex is in a tiny cycle. 
	
	\item[(v)] $\pp(X)$ is a graded poset with rank function $\rho:\pp(X)\to \mathbb Z$ which assigns to each
	$X$-cactus $\N$ its {\em rank} $\rho(\N)$ given by $\rank (\N)=0$ in case $\N$ is the trivial $X$-cactus
	and, otherwise, by
	\label{p:part:rank} %$\rank$:
	% in~\eqref{eq:rank}.
	\begin{equation}
	%	\label{eq:rank}
	\rank(\N)=\sum_{\sigma\in \C(\N)} \chi(\sigma),
	\nonumber
	\end{equation}
	where  $\chi(\sigma)= \max\{1,|\underline{\sigma}|-2\}$.
	%Given a circular partition $\sigma$ of $X$, let $|\sigma|$ be the size of $\sigma$, that 
	%is, the number of parts contained in $\sigma$.
	
	%Note that $\rank(\sigma)=1$ if and only if the size of $\sigma$ 
	%is either two (e.g. $\sigma$ is  a split) or three. 
	
	\item[(vi)] If $\N \in \pp(X)$, then $0\le \rank(\N) \le 3|X|-5$. 
	Moreover, $\rank(\N)=3|X|-5$ if and only if $\N$ is 
	a binary phylogenetic $X$-cactus in which 
	every internal vertex is contained in some tiny cycle.	
		\label{p:part:bound}

\end{itemize}
\end{thm}

\begin{proof}
(i): Suppose $\N, \N',\N''\in\pp(X)$. Then, clearly, $\N\le\N$ since we can 
take the empty sequence of contractions. Hence,  $\le$ is reflexive. 
In addition, since a contraction applied to a non-trivial  $X$-cactus reduces the number of  
its edges by at least one, it follows that $\N\le\N'$ and $\N'\le\N$ 
together imply $\N=\N'$. So $\le$ is  antisymmetric.
Finally, if $\N\le\N'$ and $\N'\le\N''$, then there is a 
sequence of contractions from $\N''$ to $\N'$, and a sequence of 
contractions from $\N'$ to $\N$, and therefore a 
sequence of contractions from $\N''$ to $\N$.  Thus $\N\le\N''$,
and so $\le$ is transitive. Thus, $\leq$ is a partial order on $\pp(X)$.

\noindent (ii): This follows immediately from~(i).

\noindent (iii): Clearly the trivial $X$-cactus is a minimal element as no 
contraction may be applied.  Any other $X$-cactus $\N$ has at least 
one edge $e$ which either induces a split 
or is contained in a cycle $C$.  If $e$ induces a split, then
an edge contraction of $\N$ on $e$ 
can be performed.  
If $e$ is in $C$, then if $C$ has four or more vertices, an edge 
contraction of $\N$ on $e$ can be performed. Otherwise $C$ is tiny 
and a triple contraction of $\N$ on $C$ can be performed.
In either case, $\N$ is not minimal.

\noindent (iv): Suppose $\N=(N,\phi)$ is a maximal element in $\pp(X)$ that is not 
of the form given in the statement of (iv).
If $\N$ has a internal vertex $v$ that is contained in $\phi(X)$ or a leaf $v$ 
with $|\phi^{-1}(v)|\geq 2$  then, for every $x \in \phi^{-1}(v)$, we add 
a new vertex $w_x$ which we label by $x$ and the edge $\{v,w_x\}$ and remove
all labels of $v$ under $\phi$. This results in an $X$-cactus $\N'$.
The $X$-cactus $\N$ can then be obtained from $\N'$ 
by performing a non-empty sequence (possibly of length 1 in case
$|\phi^{-1}(v)|=1$) 
of edge contractions of $\N'$ on the edges $\{v,w_x\}$ (in any order).
So, we may assume that $\N$ is a phylogenetic $X$-cactus.

Now, if an (unlabelled) internal vertex in $\N$ has degree four or more, 
then that vertex can be ``popped" into two new vertices by inserting an
edge $e$. Since this is clearly the reverse of an edge contraction of
the resulting $X$-cactus on $e$, we may assume that every internal vertex in $\N$ has degree 3.

Finally, in case there is an internal vertex $v$ in $\N$ of degree three that is
not contained in a cycle, then we may replace $v$ by a 3-cycle, i.e. perform the 
reverse of a triple contraction of $\N$ on $v$. Statement (iv) now follows.

\noindent (v):  Since $\pp(X)$ is finite, for any $\N, \N' \in \pp(X)$
with $\N > \N'$ there is a sequence
$\N = \N_1, \N_2, \dots ,\N_m = \N'$ of elements in $\pp(X)$, such that 
$\N_{i+1}$ is a coatom of $\N_i$ for all $1 \le i \le m-1$. Hence it suffices to 
prove that if $\N_1,\N_2\in\pp(X)$ are such that $\N_1$ is a coatom of $\N_2$,
then $\rank(\N_2)=\rank(\N_1)+1$. 

In view of Statement~(ii), $\N_1$ is obtained from $\N_2$ by one contraction, and
so we shall make a case analysis according to the type of 
contraction employed.   First, suppose that that contraction is a triple contraction of $\N_2$ on 
a (necessarily tiny) cycle $C$. Let $\sigma$ denote the circular partition induced by $C$. 
Then $|\underline{\sigma}|=3$, $\sigma\not \in \C(\N_1)$ and  $\C(\N_2) =\C(\N_1) \cup \{\sigma\}$. 
This implies $\rank(\N_2)=\rank(\N_1)+1$. 

Now, suppose that the contraction is an
edge contraction of $\N$ on some edge $e$. 
There are two subcases to consider. If $e$ is a cut edge, then 
let $\sigma$ denote the split of $X$ induced by $e$. 
Then $|\underline{\sigma}|=2$, $\sigma\not \in \C(\N_1)$ and  
$\C(\N_2) =\C(\N_1) \cup \{\sigma\}$. Thus, $\rank(\N_2)=\rank(\N_1)+1$ follows again. 
Otherwise, $e$ is contained in a cycle $C$ of size at least four.
Let $C'$ be the cycle in $\N_1$ obtained by contracting $e$.
Denote the circular partitions induced by $C$ and $C'$ by $\sigma$ and $\sigma'$, 
respectively. Then $|\underline{\sigma}|=|\underline{\sigma'}|+1$ and 
hence $\chi(\sigma)=\chi(\sigma')+1$.  
So $\rank(\N_2)-\rank(\N_1)=\chi(\sigma)-\chi(\sigma')=1$, which completes 
the proof of Statement~(v).

\noindent (vi):  Let $n=|X|$. 
Since a binary phylogenetic $X$-cactus  in which 
every internal vertex is contained in some tiny cycle 
has rank $3n-5$, we need to show that $\rho(\N)\le 3n-5$ 
holds for every $X$-cactus $\N=(N,\phi)$ in $\pp(X)$ and also that the equality holds only if every cycle in $\N$ is tiny.
If $\N$ is the trivial $X$-cactus then $\rho(\N)=0$ and the stated inequality follows. So assume that $\N$ is not trivial. 
If $n=2$ then $\rho(\N)=1$ as the unique element in $\C(\N)$ is the split induced by the sole edge of $\N$. The stated inequality follows again.

So assume $n\geq 3$.  We use induction on $n$. 
The base case $n=3$ is a
straightforward consequence  of the fact that the maximal element in $\pp(X)$ is the $X$-cactus for which every internal vertex is contained in its unique tiny cycle. 

Asume that $n>3$. Then the stated inequality holds for any 
$X$-cactus $\N'$ in  $\pp(X')$ with $3\le |X'|<n$. For the 
induction step, assume that $\N=(N,\phi)$ is an $X$-cactus in $\pp(X)$ 
with $|X|=n$. Without loss of generality, we may assume  
that $\N$ is a maximal element in $\pp(X)$ under $\leq$. 
By Statement~(iv), $\N$ is a binary phylogenetic $X$-cactus and every internal vertex is contained in some cycle. First we shall show that $\rank(\N)\le 3n-5$. 
%Without loss of generality, we may assume that $\N$ is a 
%maximal element in $\pp(X)$ under $\leq$. 
%By Statement~(iv), $\N$ is a binary phylogenetic $X$-cactus and %every 
%internal vertex is contained in some cycle. 
Let $x \in X$ and put $X'=X-\{x\}$. 
Then there exists a leaf $v_x$ in $\N$ such that $x\in\phi^{-1}(v_x)$.
 Let $u_x$ be a vertex in $\N$ adjacent with $v_x$. 
Then $e=\{v_x,u_x\}$ is a cut edge in $\N$. Let $\sigma_x=x|X'$ 
be the split induced by $e$. Note that $u_x$ is an internal
vertex of $\N$ and hence contained in some cycle $C_x$ 
and unlabelled under $\phi$ because $\N$ is a phylogenetic $X$-cactus. 
Denote the circular partition induced by $C_x$ by $\sigma$ and 
the  two vertices in $C_x$ adjacent to $u_x$ by $u_1$ and $u_2$.
Note that $u_1$ and $u_2$ must exist as $\N$ is not a simple graph. 
Note that $\sigma$ and $\sigma_x$ are the only two partitions in $\C(\N)$ that contain $\{x\}$ as a part. We now have two subcases to consider:

First, suppose $C_x$ contains at least four vertices, i.e., $u_1$ and $u_2$ are not adjacent.  
Deleting the three edges incident with $u_x$ and also the leaf $v_x$
and adding an 
edge between $u_1$ and $u_2$ results in a network $\N'$ on $X'$, 
in which $u_1$ and $u_2$ are contained in a cycle $C'$. 
Let $\sigma'$ be the circular partition induced by $C'$. 
Let $\rho':\pp(X')\to \mathbb Z$ denote the rank function 
for the graded poset $\pp(X')$ which we define analogously to
$\rho$ but with $\rho$ replaced by $\rho'$ and $\chi$ replaced by $\chi'$.
Then we have $\chi'(\sigma')=\chi(\sigma)-1$. Thus,
$\rho(\N)-\rho'(\N')=(\chi(\sigma)-\chi'(\sigma'))+\chi(\sigma_x)=2$. 
Together with the induction hypothesis, it follows that $\rho(\N) =\rho'(\N')+2\le (3n-8)+2=3n-6$ thereby 
completing the induction step. 

Now suppose that $C_x$ is tiny and contains precisely the three vertices $u_x$, $u_1$, and $u_2$. For $i=1,2$, let $v_i$ be the vertex 
adjacent to $u_i$ and not contained in $C_x$. Denote the split 
induced by the cut edge $e_i=\{u_i,v_i\}$ by $\sigma_i$.  
Deleting $C_x$, the leaf $v_x$ and the three edges
$e$, $\{u_1,v_1\}$ and $\{u_2,v_2\}$ and 
adding an edge between $v_1$ and $v_2$ results in an $X'$-cactus
$\N'$. Note that $\{v_1,v_2\}$ is a cut 
edge in $\N'$. Denote the split it induces by $\sigma'$. Note 
that $\sigma'$ can be obtained by deleting $x$ from 
either  $\sigma_1$ or $\sigma_2$. Therefore, we have 
$$
\rho(\N)-\rho'(\N')=\chi(\sigma)+\chi(\sigma_x)+\chi(\sigma_1)+\chi(\sigma_2)-\chi'(\sigma')=3
$$
since each term in the middle sum equates to $1$.
By the induction hypothesis, it follows that $\rho(\N) =\rho'(\N')+3\le (3n-8)+3=3n-5$. 
Furthermore, the equality holds only if $\N'$ is an $X'$-cactus
in which every cycle is tiny. 
By construction, every cycle in $\N'$ is tiny if and only if 
every cycle in $\N$ is 
tiny. This completes the proof of the induction step 
for this subcase too and therefore the proof of Statement~(vi).
\end{proof}

\begin{rem}
	Note that the poset $(\mathcal T(X),\leq)$ is pure 
	(i.\,e.\,bounded and all maximal chains have 
	the same length), but the poset $(\pp(X),\le)$ 
	does not have this property.
\end{rem}

\section{A characterization of the $X$-cactus ordering}\label{sec:characterize}
\label{sec:char}

In this section, we present a characterization of the partial order $\le$ on $\pp(X)$
based on the collection of circular partitions associated to an
$X$-cactus in $\pp(X)$.  To this end, given  
two non-empty sets of circular partitions $\C$ and $\C'$ in $\C(X)$, 
a map $\lift: \C \to \C'$ is called a {\em domination map} if $\lift$ 
is injective and $ \sigma \preceq \lift(\sigma)$ holds for each $\sigma\in \C$. 
%A domination map is called {\em canonical} if $\bn(\lift(C))\le \bn(\lift'(C))$ holds for each domination map $\lift'$ from $\C$ to $\C'$. 
We say that $\C$ is {\em dominated} by $\C'$, denoted by $\C \dom \C'$, if 
there exists a domination map from $\C$ to $\C'$. 
We use the convention that the empty set $\emptyset$ is dominated 
by any set $\C$ of circular partitions. In this case we also
put $\emptyset\dom\C$. 

The main result of this section can now be stated as follows:

\begin{thm}
	\label{thm:poset:net:partition}
	$\N \le \N'$ holds for two $X$-cactuses $\N$ and $\N'$ if and only if $\C(\N) \dom \C(\N')$.
	%$\bps(\N)\subseteq\bps(\N')$ and $\pps(\N) \dom \pps(\N')$.  
\end{thm}

The proof of Theorem~\ref{thm:poset:net:partition} is presented later on in this 
section and relies on a number of intermediate results. We start by presenting an observation 
concerning the poset $(\C(X),\preceq)$.

\begin{lem}
	\label{lem:refinement:compatible}
	Suppose that $\sigma_1,\sigma_2$ and $\sigma'_2$ are three circular partitions in $\C(X)$
	such that $\sigma_1 \preceq \sigma_2$ and $\sigma_2$ is compatible with $\sigma'_2$. 
	Then $ \sigma_1$ is compatible with $\sigma'_2$.
	Moreover, if $\sigma'_1$ is a circular partition with $\sigma'_1\preceq \sigma'_2$, 
	then $\sigma_1$ is compatible with $\sigma'_1$.
\end{lem}

\begin{proof}
	Since $\sigma_2$ and $\sigma'_2$ are compatible, there exists a part $X_2$ in $\sigma_2$ 
	and a part $X'_2$ in $\sigma'_2$ with $X_2\cup X'_2=X$. 
	Because $\sigma_1\preceq \sigma_2$, there exists a part $X_1$ in $\sigma_1$ 
	with $X_2\subseteq X_1$.  This implies $X_1\cup X'_2=X$, and hence  $ \sigma_1$ is 
	compatible with $\sigma'_2$.
	Furthermore, since $\sigma'_1\preceq \sigma'_2$, there 
	exists a part $X'_1$ in $\sigma'_1$ with $X'_2\subseteq X'_1$. 
	Therefore  $X_1\cup X'_1=X$. Hence, $\sigma_1$ and $\sigma'_1$ are compatible. 
\end{proof}

\begin{lem}
	\label{lem:dom:basic}
Suppose that $\C_1$, $\C_2$ and $\C_3$ are three sets of circular partitions in $\C(X)$. 
\begin{itemize}
	\item[(i)] If $\C_1\subseteq \C_2$, then $\C_1\dom \C_2$. 
	\item[(ii)] $\C_1\dom \C_2$ if and only if $\C_1\cap \bps(X) \subseteq \C_2\cap \bps(X)$ and $\C_1\cap \pps(X) \dom \C_2\cap \pps(X)$. 
	\item[(iii)] If $\C_1\dom \C_2$ and $\C_2 \dom \C_3$, then $\C_1\dom \C_3$.
\end{itemize}	
\end{lem}	
 
 \begin{proof}
We assume $\C_1 \neq \emptyset$ as the lemma clearly holds otherwise.
 	
\noindent (i) Note that the map $\lift: \C_1 \to \C_2$ defined as  $\lift(\sigma)=\sigma$ is a domination map. 
%holds for each partition $\sigma$ in $\C_1$.

\noindent (ii) It is straightforward to show that the
	statement holds if $\C_b(X)\cap \C_1=\emptyset$ or
	if $\C_p(X)\cap \C_1=\emptyset$.
	So assume that $\C_b(X)\cap \C_1\not=\emptyset$ and that
	$\C_p(X)\cap \C_1\not=\emptyset$.
	 Assume first that $\C_1\dom \C_2$ and consider a domination map $\lift: \C_1 \to \C_2$. Let $\sigma\in \C_1$.
Then $\sigma \in \bps(X)$ if and only if $\lift(\sigma)$ is 
contained in $\bps(X)$. This implies that the restriction of $\lift$ to $\C_1\cap \bps(X)$ 
is a domination map from $\C_1\cap \bps(X)$ to  $\C_2\cap \bps(X)$. 
Thus, we have $\C_1\cap \bps(X) \subseteq \C_2\cap \bps(X)$. Furthermore, 
since $\sigma\in \pps(X)$ if and only if $\lift(\sigma)\in \pps(X)$, 
it follows that $\C_1\cap \pps(X) \dom \C_2\cap \pps(X)$.

Conversely, assume that $\C_1\cap \bps(X) \subseteq \C_2\cap \bps(X)$ and that
$\C_1\cap \pps(X) \dom \C_2\cap \pps(X)$. By Part~(i) 
of the lemma, there exists a domination map $\lift_b$  from $\C_1\cap \bps(X)$ to $\C_2\cap \bps(X)$. Let 
$\lift_p$ be a domination map from $\C_1\cap \pps(X)$ to $\C_2\cap \pps(X)$. 
Now consider the map $\lift: \C_1 \to \C_2$ with $\lift(\sigma)=\lift_b(\sigma)$ if $\sigma\in \C_1\cap \bps(X)$ 
and $\lift(\sigma)=\lift_p(\sigma)$ if $\sigma\in \C_1\cap \pps(X)$. 
Then $\lift$ is a domination map. Hence, $\C_1\dom \C_2$.

\noindent (iii)  Fix a domination map $\lift_1: \C_1 \to \C_2$ and a 
domination map $\lift_2: \C_2 \to \C_3$. Then the map $\lift: \C_1\to \C_3$  
defined by  $\lift(\sigma)=\lift_2(\lift_1(\sigma))$ for all
$\sigma\in \C_1$ is a domination map. Indeed, 
$\lift$ is injective since both $\lift_1$ and $\lift_2$ are injective, and, 
$\sigma\preceq \lift_1(\sigma)$ and $ \lift_1(\sigma)\preceq \lift_2(\lift_1(\sigma))$  
imply $\sigma\preceq \lift(\sigma)$. Hence, $\C_1\dom \C_3$.
 \end{proof}
 
Given any two sets of circular partitions $\C_1$ and $\C_2$ in
$\C(X)$ with  $\C_1 \dom \C_2$, 
there could in general be several domination maps from $\C_1$ to $\C_2$. 
For instance,  let $\sigma_1=\big[ 12|3|45|6 \big]$ and $\sigma_2=\big[ 1|2|3|45|6 \big]$,  
and consider $\C_1=\{\sigma_1\}$ and $\C_2=\{\sigma_1,\sigma_2\}$. 
Then $\C_1\dom \C_2$ and there are two domination maps from $\C_1$ to $\C_2$: 
one maps $\sigma_1$ to $\sigma_1$, and the other maps $\sigma_1$ to $\sigma_2$. 
However, the following lemma shows that when $\C_2$ is compatible, the domination map is unique. 

\begin{lem}
\label{lem:dom:compatible}
Suppose that $\C_1$ and $\C_2$ are two non-empty sets of circular partitions in $\C(X)$ and that $\C_2$ is compatible. 
Then for each circular partition $\sigma_1\in\C_1$, there exists at 
most one circular partition $\sigma_2\in \C_2$ with $\sigma_1 \preceq \sigma_2$. 
Moreover, $\C_1 \dom \C_2$ holds if and only if there exists a unique domination map from $\C_1$ to $\C_2$.
\end{lem}

\begin{proof}
Suppose that $\sigma_1\in\C_1$  and that  $\sigma_2$ and $\sigma'_2$ are two circular partitions in $\C_2$
with $\sigma_1 \preceq \sigma_2$ and $\sigma_1\preceq \sigma_2'$. We shall show that $\sigma_2=\sigma'_2$. 
This clearly holds if $\sigma_1$ is a split because in this case we have $\sigma_2=\sigma_1=\sigma'_2$. 
So, assume for contradiction that $\sigma_1$ is proper, that is, $|\underline{\sigma_1}|>2$, and 
that $\sigma_2\not =\sigma'_2$. Since $\C_2$ is compatible, 
there exists a part $X_2$ in $\sigma_2$ and $X'_2$ in $\sigma'_2$ 
such that $X_2\cup X_2'=X$.  On the other hand, $\sigma_1\preceq \sigma_2$ 
implies that there exists a part $X_1$ in $\sigma_1$ with $X_2 \subseteq X_1$. 
Similarly, since $\sigma_1\preceq \sigma_2'$ it follows that there exists a part 
$X'_1$ in $\sigma_1$ with $X'_2 \subseteq X'_1$. However, this implies $X_1\cup X'_1=X$, 
a contradiction to the fact that $\sigma_1$ is proper. 
Thus $\sigma_2=\sigma'_2$, completing the proof of the first part of the lemma. 
 
To establish the second part of the lemma, it clearly suffices
to show that if $\C_1\dom\C_2$ holds then there must exist a unique
domination map form $\C_1$ to $\C_2$. So
assume $\C_1\dom\C_2$ 
and that there exist two distinct domination maps 
$\lift$ and $\lift'$ from $\C_1$ to $\C_2$. Then there exists a 
circular partition $\sigma$ in $\C_1$ with $\lift(\sigma)\not= \lift'(\sigma)$. 
Since $\lift$ and $\lift'$ are domination maps we have $ \sigma\preceq \lift(\sigma)$ 
and $ \sigma\preceq \lift'(\sigma)$, a contradiction to  
the first part of the lemma. 
\end{proof}

\noindent With these results in hand, we now prove Theorem~\ref{thm:poset:net:partition}:

\begin{proof}
First assume that $\N$ and $\N'$ are 
two $X$-cactuses for which $\N\leq \N'$ holds.
Without loss of generality, we may assume that $\N$ is obtained 
from $\N'$ by one contraction. Furthermore, we may assume that
	$\N$  is not the trivial $X$-cactus since in this case the
	theorem holds in view of Theorem~\ref{t:bijection.QX.CN}. %and Remark~\ref{rem:convention}.
By Lemma~\ref{lem:dom:basic}(ii) 
it suffices to show that $\bps(\N)\subseteq\bps(\N')$ and $\pps(\N) \dom \pps(\N')$.  
To this end, we consider the following possible three cases:

First, assume that $\N'$ is obtained from $\N$ by an edges
contraction of $\N$ on a cut edge. 
Then $\bps(\N)\subset \bps(\N')$. Moreover, $\pps(\N) = \pps(\N')$, and so
$\pps(\N) \dom \pps(\N')$ in view of Lemma~\ref{lem:dom:basic}(i).

Next, assume that  $\N'$ is obtained from $\N$ by a triple contraction of $\N$ on a tiny cycle $C$.   
Then $\bps(\N)=\bps(\N')$ and $\pps(\N) \subset \pps(\N')$, and so 
$\pps(\N) \dom \pps(\N')$ by Lemma~\ref{lem:dom:basic}(i).

Finally, assume that $\N'$ is obtained from $\N$ by an edge  contraction of $\N$ on an edge $e$ 
in a cycle $C$ that is not tiny. In particular, $C$ is contracted to 
a cycle $C'$ in $\N'$ with one less edge.  Then $\bps(\N)=\bps(\N')$. 
Denote the circular partition induced by $C$ in $\N$  
and $C'$ in $\N'$ by $\sigma$ and $\sigma'$, respectively. 
Then $\pps(\N')=(\pps(\N)-\{\sigma\})\cup \{\sigma'\}$, from which it is straightforward 
to verify that $\pps(\N) \dom \pps(\N')$ by noting that $\sigma' \preceq \sigma$ and 
using Theorem~\ref{t:bijection.QX.CN} and Lemma~\ref{lem:dom:compatible}. 

Conversely, suppose that $\N$ and $\N'$ are two $X$-cactuses with $\C(\N) \dom \C(\N')$.
By  Lemma~\ref{lem:dom:basic}(ii),  $\bps(\N)\subseteq\bps(\N')$ and $\pps(\N) \dom \pps(\N')$.  
If $\C(\N)=\emptyset$ then $\N=\N(\C(\N))$ is the trivial $X$-cactus and so $\N\leq \N'$ must hold. 

So assume that $\C(\N)\not=\emptyset$. Since $\pps(\N')$ is compatible, by Lemma~\ref{lem:dom:compatible} there 
is a unique  domination map from $\pps(\N)$ to $\pps(\N')$. Let $\C^* \subseteq \pps(\N')$ be
the image of this map. Since a domination map is injective, $|\C^*|=|\pps(\N)|$. 

Now, each split $\sigma$ in $\bps(\N')-\bps(\N)$ corresponds to a cut edge 
in $\N'$ that induces $\sigma$. Consider the network $\N_1$ 
obtained from $\N'$ by performing an edge contraction of
 $\N'$ on all cut edges that induce  some split in $\bps(\N')-\bps(\N)$.
Then $\bps(\N_1)=\bps(\N)$ and $\pps(\N_1)=\pps(\N')$.

Next, for each proper partition $\sigma$ in $\pps(\N_1)-\C^*$,
there is a cycle $C_{\sigma}$ in $\N_1$ that induces $\sigma$. Consider the 
network $\N_2$ obtained from $\N_1$ by performing a triple contraction
of $\N_1$ on all cycles in $\N_1$ that correspond to some proper 
partition in $\pps(\N_1)-\C^*$ (i.e., for every $\sigma \in \pps(\N_1)-\C^*$, first apply a (possibly empty) sequence 
of edge contractions to covert $C_{\sigma}$ into a tiny cycle, 
and then apply a triple contraction on the resulting tiny cycle). 
Then $\bps(\N_2)=\bps(\N)$ and $\pps(\N_2)=\C^*$.

Finally, for each proper partition $\sigma^*$ in $\C^*$, there is a cycle $C_{\sigma^*}$ in $\N_2$ 
that induces $\sigma^*$. Given such a 
partition $\sigma^*$, let $\sigma$ be the unique circular 
partition in $\pps(\N)$ with $\sigma \preceq \sigma^*$. 
For every $C_{\sigma^*}$ in $\N_2$
perform a (possibly empty) series of edge contractions on $\N_2$ 
so that in the resulting $X$-cactus $\N_3$ there exists a cycle
that induces $\sigma$. Then 
$\bps(\N_3)=\bps(\N)$ and $\pps(\N_3)=\pps(\N)$
because $|\C^*|=|\pps(\N)|$. By Theorem~\ref{t:bijection.QX.CN},
$\N_3$ is isomorphic to $\N$.  Since $\N_3 \leq \N'$ by construction, $\N \leq \N'$ follows.
\end{proof}
 
 We end this section by giving two consequences of Theorem~\ref{thm:poset:net:partition}. 
 First, recall that, if $(S,\le)$ and $(S',\preceq)$ 
 are arbitrary posets, then a map $f:S \to S'$ is an embedding of
 $(S,\le)$ into $(S',\preceq)$ if, for all $s,s' \in S$,  
 $s \le s'$ if and only if $f(s) \preceq f(s')$ \cite[p.436]{trotter1995partially}.
 Now, we have two natural maps $i:\TT(X) \to \pp(X)$ and $i': \C(X) \to \pp(X)$.
 The first is the inclusion map, and the second is given
 by taking a circular partition $[A_1|\dots|A_k]$ of $X$ to the 
 $X$-cactus which, if $k\geq 3$, is a cycle of length $k$ with each vertex labelled
 by $A_i$ so that the circular ordering of the $A_i$'s is preserved and, if $k=2$, is a cut edge.
 
 \begin{cor}
 	The maps $i$ and $i'$ are both poset embeddings.
 \end{cor}
 \begin{proof}
 	First consider the map $i$. Suppose $\TT_1,\TT_2 \in \TT(X)$. Then 
 	$\TT_1\le \TT_2$ in $\TT(X)$ if and only if $\C(\TT_1)\subseteq \C(\TT_2)$ holds. 
 	Since both $\C(\TT_1)$ and $\C(\TT_2)$ are contained in $\C_b(X)$, by Lemma~\ref{lem:dom:basic}(ii), we have 
 	$\C(\TT_1)\subseteq \C(\TT_2)$ if and only if $\C(\TT_1)\dom \C(\TT_2)$. Together with
 	Theorem~\ref{thm:poset:net:partition}, it follows that  $\TT_1\le \TT_2$ in $\TT(X)$ if and only 
 	if $i(\TT_1)\le i(\TT_2)$ in $\pp(X)$.
 	
 	Now consider the map $i'$. Suppose $\sigma_1,\sigma_2 \in \C(X)$. 
 	Since $\sigma_1\preceq \sigma_2$ if and only if $\{\sigma_1\}\dom \{\sigma_2\}$, 
 	by Theorem~\ref{thm:poset:net:partition} it follows that $\sigma_1\preceq \sigma_2$ 
 	if and only if $i'(\sigma_1)=\N(\{\sigma_1\})\le \N(\{\sigma_2\})=i'(\sigma_2)$.
 \end{proof}
 
 Finally, we state a useful observation about upper and lower bounds in $(\pp(X),\le)$, 
 which is a straightforward consequence of Lemma~\ref{lem:dom:basic} and Theorem~\ref{thm:poset:net:partition}. 
 
 \begin{cor}
 	\label{cor:bound:char}
 	Let $\pp=\{\N_1,\dots,\N_m\}\subseteq\pp(X)$, $m\geq 1$, be a collection of $X$-cactuses. Then the following two statements hold.
 	\begin{itemize}
 		\item[(i)] An $X$-cactus $\N$ is an upper bound of $\pp$ if and only if $ \bigcup_{i=1}^m\bps(\N_i)  \subseteq \bps(\N)$ and, for all $1\le i \le m$,  $\pps(\N_i) \dom \pps(\N)$ holds.  
 		\item[(ii)] An
 		$X$-cactus $\N$ is a lower bound of 
 		$\pp$ if and only if $\bps(\N)\subseteq \bigcap_{i=1}^m\bps(\N_i)$ and, for all $1\le i \le m$,
 		$\pps(\N) \dom \pps(\N_i)$ holds.  
 	\end{itemize}
 \end{cor}

 \section{Upper bounds}\label{sec:upper}
 
\begin{figure}[ht]
 	\begin{center}
 	\includegraphics[width=0.8\textwidth]{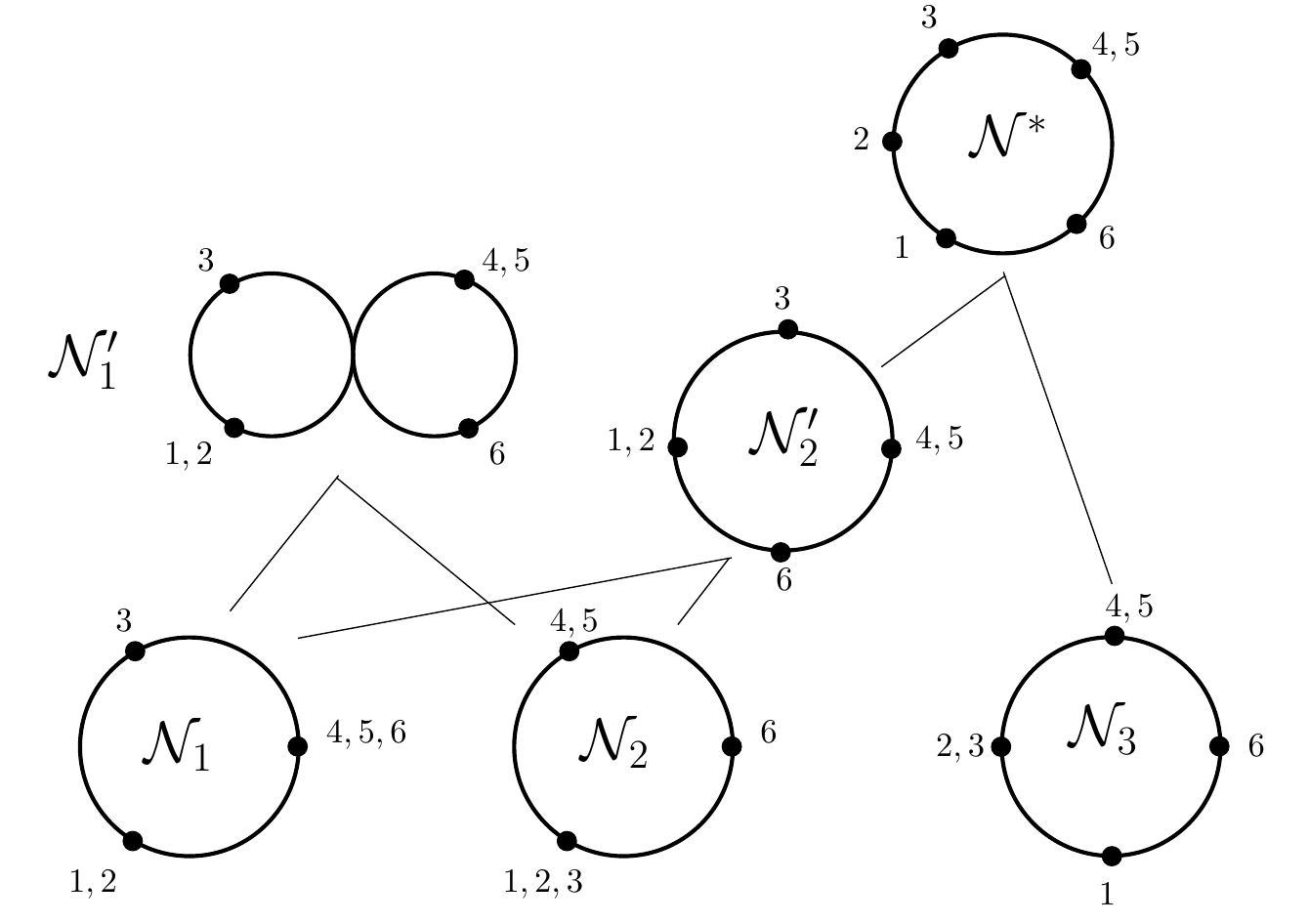}
 	\caption{The Hasse diagram of six $X$-cactuses $\N_1,\N_2,\N_3,\N'_1,\N'_2,\N^*$ where $X=\{1,2,\ldots, 6\}$. 
 		The set $\{\N_1,\N_2\}$ has two least upper bounds, $\N'_1$ and $\N_2'$,
 		whereas the set $\{\N_1,\N_2,\N_3\}$ has a unique least upper bound $\N^*$. For ease of readability, wee have omitted the brackets from the set that labels a vertex.  
 	} 
 	\label{f:non-unique.ub}
 	\end{center}
 \end{figure}

 In  Corollary~\ref{cor:bound:char}, we gave 
 a characterization for when a set $\pp \subseteq \pp(X)$ of $X$-cactuses 
 has an upper bound in  $(\pp(X),\le)$. 
 In this section, we present two further results concerning upper bounds in 
 $(\pp(X),\le)$. Upper bounds are of interest
 since, if they exist, they can be thought of as ``supernetworks'' for 
 collections of networks. Note that the behaviour of upper bounds in $(\pp(X),\le)$ is 
 more complicated than it is for the poset $(\mathcal T(X),\leq)$ of $X$-trees.
 For example, as shown in \cite[Theorem 3.3.3]{semple2003phylogenetics},
 in case a set $\mathcal T$ of $X$-trees has an upper bound in $(\mathcal T(X),\le)$, then  
 it has a unique least upper bound. However, this is not the case for $(\pp(X),\le)$
 (e.g. Figure~\ref{f:non-unique.ub}).

Our first result gives an insight on the number of 
circular partitions contained in an upper bound.

 \begin{thm}
 	\label{thm:ub:size}
 	If $\N$ is a least upper bound for a collection 
 	$\{\N_1,\dots,\N_m\}\subseteq\pp(X)$ of $X$-cactuses, some $m\geq 1$, then 
 	\[\left|\C(\N)\right|\le\big|\bigcup_{1\le i \le m }\C(\N_i)~\big|.
 	%\le \sum_{i=1}^m |\C(\N_i)|.
 	\]
 \end{thm}
 \begin{proof}
 	%Since the second inequality in the theorem clearly holds, it suffices to establish the first one. To this end, 
 	Since $\N$ is an upper bound for $\{\N_1,\ldots, \N_2\}$, by
 	Corollary~\ref{cor:bound:char}, for each $1\le i \le m$, 
 	there is a subset $\C^i_b(\N)$ of $\C_b(\N)$ so that 
 	$\C^i_b(\N)=\C_b(\N_i)$	holds. Furthermore, 
 	we can take a minimal subset $\C^i_p(\N)$ of $\C_p(\N)$ so 
 	that $\pps(\N_i) \dom \C^i_p(\N)$ holds, where 
 	minimality implies $|\pps(\N_i)| =|\C^i_p(\N)|$. 
 	We consider the splits and proper partitions separately.
 	
 	First, we claim that
 	\begin{equation}
 	\label{eq:pf:ub:size:1}
 	\C_b(\N)	=\bigcup_{1\le i \le m} \C^i_b(\N)=\bigcup_{1\le i \le m} \C_b(\N_i).
 	\end{equation}
 	The right equality holds by definition. To see that
 	the left one holds, note first that by definition 
 	$\bigcup_{1\le i \le m} \C^i_b(\N)\subseteq \C_b(\N)$. 
 	To see that equality holds,  assume for contradiction that this is not the case, i.\,e.\,that there exists a split $\sigma$
 	in $\C_b(\N) - \bigcup_{1\le i \le m} \C^i_b(\N)$. Let 
 	$\N^*$ be the $X$-cactus obtained from $\N$ by contracting the edge in $\N$
 	corresponding to $\sigma$. Then by Corollary~\ref{cor:bound:char}, 
 	$\N^*$ is an upper bound of $\{\N_1,\dots,\N_m\}$ with 
 	$\N^*< \N$, a contradiction. 
 	
 	We next claim
 	\begin{equation}
 	\label{eq:pf:ub:size:2}
 	|\C_p(\N)|	= \Big| \bigcup_{1\le i \le m} \C^i_p(\N) \Big| \le \Big| \bigcup_{1\le i \le m} \C_p(\N_i) \Big|.
 	\end{equation}
 	The left equality can be shown to 
 hold using a similar argument to the one used to prove  Equality~\eqref{eq:pf:ub:size:1}. 
 	To see that the right inequality holds, 
 	for each circular partition $\sigma\in \bigcup_{1\le i \le m} \C^i_p(\N)$, 
 	let $t(\sigma)$ be the smallest index in $\{1,\dots,m\}$ 
 	(subject to some ordering) 	such that $\sigma$ is 
 	contained in $\C^{t(\sigma)}_p(\N)$. By construction, there exists a unique 
 	circular partition  in $\C_p(\N_{t(\sigma)})$, denoted by $f(\sigma)$, 
 	such that $\sigma\preceq f(\sigma)$ holds. Since the map 
 	$f: \bigcup_{1\le i \le m} \C^i_p(\N) \to \bigcup_{1\le i \le m} \C_p(\N_i)$ 
 	associating each $\sigma\in\bigcup_{1\le i \le m} \C^i_p(\N)$ to $f(\sigma) \in\bigcup_{1\le i \le m} \C_p(\N_i)$ 
 	is injective, the right inequality follows. 
 	
 	The theorem follows now from Equalities ~\eqref{eq:pf:ub:size:1} and~\eqref{eq:pf:ub:size:2} 
 	since $\C(\N)$ is a disjoint union of $\C_b(\N)$ and $\C_p(\N)$, 
 	and $\bigcup_{1\le i \le m} \C(\N_i)$ is a disjoint union of 
 	$\bigcup_{1\le i \le m} \C_b(\N_i)$ and $\bigcup_{1\le i \le m} \C_p(\N_i)$. 
 \end{proof}
 
 In our second result, we give an alternative characterization to Corollary~\ref{cor:bound:char}
 for when the upper bound for two $X$-cactuses exists (Theorem~\ref{thm:up}), which
 gives some more structural insights into determining whether this is the case or not. 
 %Extending \cite[Definition 3.3.1]{semple2003phylogenetics} for $X$-trees, we shall say 
 %that a subset $\pp \subseteq \pp(X)$ of $X$-cactuses is {\em %compatible} if it has an 
 %upper bound under the ordering $\le$. 
 
 Now, for distinct $X$-cactuses $\N_1,\N_2\in\pp(X)$ 
 we define the {\em incompatibility graph} $\IG(\N_1,\N_2)=(W,F)$ to be the
graph with vertex set $W=\C(\N_1)\cup\C(\N_2)$
 and edge set  $F$ consisting of all pairs 
 $\{\sigma_1,\sigma_2\}$  of distinct circular partitions in $W$ 
 such that $\sigma_1$ and $\sigma_2$ are incompatible. 
 A {\em resolution} of $(W,F)$ is an injective map  $\lambda: F \to \C(X)$  
 such that   $\lambda(\{\sigma_1,\sigma_2\})$ is an upper bound of  $\sigma_1$ and $\sigma_2$ 
 in  $(\C(X),\preceq)$,  for every edge $\{\sigma_1,\sigma_2\} \in F$. 
 Such a resolution is called {\em minimal} if $\lambda(\{\sigma_1,\sigma_2\})$ is 
 a least upper bound of $\sigma_1$ and $\sigma_2$ in $(\C(X),\preceq)$ 
 for each edge $\{\sigma_1,\sigma_2\}$ in $F$. Note that  
 if $F=\emptyset$, then we use the convention that 
 the empty function $\lambda:F\to \C(X)$ with $\lambda(F)=\emptyset$ 
 is the (necessarily unique) minimal resolution of $(W,F)$.  
%The set of isolated vertices (i.e. degree-$0$ vertices) is denoted by $V_0(G)$. 

 \begin{thm} \label{thm:up}
 	Suppose $\N_1$ and $\N_2$ are two distinct
 	$X$-cactuses. Then the following statements are equivalent.
 	\begin{enumerate}
\item[(i)] 	
 	$\N_1$ and $\N_2$ have an upper bound under $\le$;  
 	\item[(ii)] the incompatibility graph $\IG(\N_1,\N_2)=(W,F)$ is a matching (i.e. 
 	every vertex has degree $0$ or $1$),
 	and there exists a resolution $\lambda$ of $(W,F)$ such that $W_0 \cup \lambda(F)$ is compatible (where
 	$W_0\subseteq W$ denotes the set of isolated vertices in $\IG(\N_1,\N_2)$).  
% 	\item[(iii)]	there exists a minimal solution $\lambda$ of $(W,F)$ such that $\N(W_0 \cup \lambda(F))$  	is a least upper bound for $\N_1$ and $\N_2$.
\end{enumerate}
Moreover, if  the incompatibility graph $(W,F)$ is a matching and
 there exists a minimal resolution $\lambda$ of $(W,F)$ such that $W_0 \cup \lambda(F)$ is compatible,  
%  $\lambda$ is a minimal resolution of $(W,F)$, 
then $\N(W_0 \cup \lambda(F))$  	is a least upper bound for $\N_1$ and $\N_2$.

%Statement~(ii) holds then there exists a minimal solution $\lambda$ of $(W,F)$ such that $\N(W_0 \cup \lambda(F))$ is a least upper bound for $\N_1$ and $\N_2$.
% If $F=\emptyset$, then $\N(W)$ is a least  upper bound for $\N_1$ and $\N_2$.
 \end{thm}

 Before proceeding with the proof, to illustrate Theorem~\ref{thm:up} consider the $X$-cactuses 
 $\N_1$ and $\N_2$ with $X=\{1,2\ldots, 9\}$ pictured
 in Figure~\ref{f:1_Fig_poset}.
 Then the  $X$-cactus depicted in Figure~\ref{f:nc-construction} is a 
 least upper bound for $\N_1$ and $\N_2$ since the incompatibility graph $\IG(\N_1,\N_2)$
 is a matching with sole edge $e=\{\beta,[678912|3|45]\}$ where $\beta=[678912|3|4|5]$ 
 and the map $\lambda$ assigning $e$ to $\beta$ is a minimal resolution.
 
 \begin{proof} 
 	For simplicity, put $\IG=\IG(\N_1,\N_2)=(W,F)$.
 	%\kh{In view of the change in the statement of the theorem, this paragraph might be redundant.} As the theorem clearly holds when $\N_1=\N_2$ or one of the these two $X$-cactus is trivial, in the following proof we may assume that $\N_1$ and $\N_2$ are distinct and neither of them is trivial.  
 	
 	\noindent $(i)\Rightarrow (ii)$:
 	%Let $\IG=\IG(\N_1,\N_2)=(W,F)$.
 	Suppose that $\N'$ is an upper bound of $\N_1$ and $\N_2$ under $\le$. Note that if 
  $F=\emptyset$,  then $\IG$ is a matching as every vertex in $\IG$ is isolated. Furthermore,   the empty function $\lambda:F\to \C(X)$ with $\lambda(F)=\emptyset$ is a resolution of $\IG$ and $W_0\cup \lambda(F)=W=\C(N_1)\cup\C(\N_2)$ is compatible. Hence we may assume that $F\not=\emptyset$.

 	By Lemma~\ref{lem:dom:compatible} and Theorem~\ref{thm:poset:net:partition}, 
   for $i=1,2$	there is a unique domination 
 	map from $\C(\N_i)$ to $\C(\N')$ which we denote by $L_i$.
 	We first claim that for each edge $e=\{\sigma_1,\sigma_2\} \in F$, 
 	with $\sigma_i\in \C(\N_i)$ for $i=1,2$, we have 
 	$\lift_1(\sigma_1)=\lift_2(\sigma_2)$. Indeed, 
 	if this were not the case, then $\lift_1(\sigma_1)$ and $\lift_2(\sigma_2)$ 
 	would be a pair of compatible circular partitions by Lemma~\ref{l:CN.compatible} 
 	(since both are contained in $\C(\N')$). Hence,
 	by the last part of Lemma~\ref{lem:refinement:compatible} and because  $\sigma_i\preceq \lift_i(\sigma_i)$ for $i=1,2$, 
 	it follows that $\sigma_1$ and $\sigma_2$ are compatible, a contradiction as $e\in F$. This proves the claim.
 	
 	We now show that $\IG$ is a 
 	matching. Suppose this were not the case. Then there exists a vertex in $W$
 	with degree two or more. 
 	Switching the index if necessary, we may assume that $\sigma_2$ 
 	is contained in $\C(\N_2)$, and that $\sigma_1$ and $\sigma_3$ 
 	are vertices in $\C(\N_1)$ that  are 
 	adjacent with $\sigma_2$. By the previous claim it follows 
 	that $\lift_1(\sigma_1)=\lift_1(\sigma_3)$, a contradiction 
 	to the fact that $\lift_1$ is injective. 
 	
 	Next, we show that there exists a resolution $\lambda:F\to \C(X)$ of $\IG$ as stated in Statement (ii). For
 	each edge $e=\{\sigma_1,\sigma_2\} \in F$, $\sigma_i\in \C(\N_i)$, $i=1,2$, 
 	we define  $\lambda(e)=\lift_1(\sigma_1)=\lift_2(\sigma_2)$
 	which is clearly well-defined  in view of the previous claim. Furthermore, $\lambda$ is injective because both $\lift_1$ and $\lift_2$ are injective. Since, by definition, 
 		$\lambda(e)$ is an upper bound for $\sigma_i$, $i=1,2$, it 
 		follows that $\lambda$ is a resolution of $\IG$.

 	It remains to show that $W_0\cup \lambda(F)$ is compatible.
 		To see this, note
 	that since $\lambda(e)$ is an upper bound of both $\sigma_1$ and $\sigma_2$ in $(\C(X),\preceq)$, it follows that  $W_0\cup \lambda(F)$ is a set of circular partitions.
 	Assume that $\sigma$ and $\sigma'$ are two distinct circular partitions in $W_0\cup \lambda(F)$. 
 	Note that these two circular partitions are clearly compatible if both of them 
 	are contained in $W_0$ or both in $\lambda(F)$ because 
 	$\lambda(F)\subseteq \C(\N')$ and $\C(\N')$ is compatible. Therefore, without loss 
 	of generality, we may assume that $\sigma\in W_0\cap \C(\N_1)$ 
 	and $\sigma'=\lambda(e)$ with $e=\{\sigma_1,\sigma_2\}\in F$ for $\sigma_1\in \N_1$ and $\sigma_2\in \N_2$.  
 	Since  	$\sigma\preceq \lift_1(\sigma)$ and noting that 
 	$\lift_1(\sigma)$ and $\lift_1(\sigma_1)=\sigma'$ are compatible (as 
 	they are two distinct circular partitions contained in $\C(\N')$), it follows by the first part of Lemma~\ref{lem:refinement:compatible} 
 	that $\sigma$ and $\sigma'$ are compatible. 
 	
 	%To see that Statement~(b) holds, assume that $F=\emptyset$. Then every vertex in $\IG$ is isolated. Thus $W=\C(N_1)\cup\C(\N_2)$ and so  $W$ is compatible.

 \noindent	$(ii)\Rightarrow(i)$: Suppose $\IG$  is a matching. 
 We assume $F \not= \emptyset$; the case $F = \emptyset$ can be established in a similar way.
Fix a resolution map $\lambda:F\to \C(X)$ as in Statement~(ii). Since $W_0\cup \lambda(F)$ is compatible, 	by Theorem~\ref{t:bijection.QX.CN}, 
 	there exists an $X$-cactus $\N^*$ such that $\C(\N^*)=W_0\cup \lambda(F)$. Consider the map $L_\lambda: \C(\N_1) \to \C(\N^*)$ defined as follows. If $\sigma\in \C(\N_1)$ 
 	is an isolated vertex in $\IG$, then $\sigma\in W_0$ and we let $L_\lambda(\sigma)=\sigma$; otherwise there exists a 
 	unique circular partition $\sigma_2$ in $\C(\N_2)$ such that $\{\sigma,\sigma_2\}$ 
 	is an edge in $\IG$ because $\IG$ is a matching. In this case, we let $L_\lambda(\sigma)=\lambda(\{\sigma,\sigma_2\})$. 
 	
 	We claim that $L_{\lambda}$ is a domination map. We
 	first show that $L_{\lambda}$ is injective. Assume for 
 		contradiction that there exist $\sigma,\sigma _1\in \C(\N_1)$
 		such that $L_{\lambda}(\sigma)=L_{\lambda}(\sigma_1)$ but 
 		$\sigma\not=\sigma_1$. By definition of $L_{\lambda}$ and the fact that $\lambda $
 		is a resolution and therefore injective, we may assume that 
 	 $\sigma\in W_0\cap \C_1(\N)$, that $\{\sigma_1,\sigma_2\}$ is an edge in $F$ with $\sigma_1\in \C(\N_1)$, and that $\sigma_2\in \C(\N_2)$. 
 	 Then $\sigma_1\preceq \lambda(\{\sigma_1,\sigma_2\})= L_{\lambda}(\sigma_1)=L_{\lambda}(\sigma)=\sigma$. Since, by assumption. $\sigma\not=\sigma_1$ it follows that
 	 	 $\sigma$ and $\sigma_1$ are two distinct incompatible circular partitions in $\C(\N_1)$, a contradiction. 
 	% 	Since a compatible subset of $\C(X)$ can not contain two distinct circular partition $\sigma$ and $\sigma'$ such that $\sigma \preceq \sigma'$ holds, 
 	 Thus, $L_\lambda$ must be injective. 	
 	%Since $L_\lambda$ is injective and $\sigma \preceq L_\lambda(\sigma)$ holds  	for each $\sigma$ in $\C(\N_1)$, 
 	Since $\sigma \preceq L_\lambda(\sigma)$ holds for all $\sigma$ in $\C(\N_1)$, it follows that $L_\lambda$ is a domination map, as claimed. 
 	
 	As $L_{\lambda}$ is a domination map, 
 	$\C(\N_1)\dom \C(\N^*)$ and so
 	$\N_1\le \N^*$ in view of Theorem~\ref{thm:poset:net:partition}.  Using a 
 	similar argument, we also have $\N_2\le \N^*$. Thus, $\N^*$ is an upper bound of  
 	$\N_1$ and $\N_2$ under $\leq$. This completes the proof of the equivalence 
 	of  Statements~(i) and (ii).
 	
 	%Next, assume that Statement~(b) holds. Then 	$F=\emptyset$. Similar arguments as in the case of the proof of Statement~(a) imply that there exists an $X$-cactus $\N^*$ for which $\C(\N^*)=\C(\N_1)\cup\C(\N_2)$ holds and that $\N^*$ is an upper bound for $\N_1$ and $\N_2$ under $\leq$. This completes the proof of the equivalence of Statements~(i) and (ii).
 	
 	To prove the remainder of the theorem,
 	assume that  the incompatibility graph $\IG=(W,F)$ is a matching and that $\lambda$ is a minimal resolution of $(W,F)$ such that $W_0 \cup \lambda(F)$ is compatible. 
 	Since a minimal resolution is in particular a resolution, our arguments in the previous two paragraphs imply that
 	 $\N^*=\N(W_0 \cup \lambda(F))$ is an upper bound for $\N_1$ and $\N_2$. It remains to show that $\N^*$ is a least upper bound. 
 	%To this end we may further assume that $F\not =\emptyset$ as the other case can be established in a similar manner. 
 	
 	Assume for contradiction that there exists an $X$-cactus $\widehat{\N}$
 	that is an upper bound for $\N_1$ and $\N_2$ such that 
 	$\widehat{\N}< \N^*$. 
 	% and there exists a map   	$\widehat{\lambda}:F\to \C(X)$ such that $\widehat{\lambda}(\{\sigma_1,\sigma_2\})$ is an upper bound  under $\preceq$ 	for all $\sigma_1$ and $\sigma_2$ with 	$\{\sigma_1,\sigma_2\}\in F$. 
 	Without loss of generality we may assume 
 	that 
 	%the sequence of contractions required to obtain $\widehat{\N}$
 	%from $\N^*$ has length two, i.e. 
 	$\widehat{\N}$ is obtained from
 	$\N^*$ by performing a single contraction. Let $\sigma \in \C(\N^*)$
 	be 	the unique circular partition contained in $\C(\N^*)$ but not in $\C(\widehat{\N})$.  Since 
 	either $\sigma\in W_0$ or $\sigma\in \lambda(F)$ holds, swapping the index of $\N_1$ and $\N_2$ 
 	if necessary we may assume that $\C(\N_1)$ contains a circular partition $\sigma_0$ such
 	that $\sigma_0\preceq \sigma$. Note that $\C(\N_1)\dom \C(\widehat{\N})$
 	in view of $\N_1\leq \widehat{\N}$ and Theorem~\ref{thm:poset:net:partition}. 
 	
 	If $\sigma$ corresponds to a cut edge $e$ of $\N^*$ (that is, $\widehat{N}$ is obtained from $\N^*$ by  an edge contraction
 	of $\N$  on the cut edge $e$ and $\sigma=\cpo(e)$) then 
 	$ \C(\N^*)-\{\sigma\}=\C(\widehat{\N})$. Furthermore, $\sigma_0=\sigma$
 	as $\sigma$ is a  split. Thus, $\sigma_0\in \C_b(\N_1)$ and $\sigma_0\not \in C_b(\widehat{\N})$, a contradiction in view of 
 	$\C(\N_1)\dom \C(\widehat{\N})$ and Lemma~\ref{lem:dom:basic}(ii).

 	If  $\sigma$ corresponds to a tiny cycle in $\N^*$ (i.e., $\widehat{\N}$ is obtained from $\N^*$ by  a triple contraction 
 	of $\N^*$ on this tiny cycle) then $\C(\N^*)-\{\sigma\}=\C(\widehat{\N})$. Furthermore, $\sigma_0=\sigma$
 	as $|\underline \sigma|=3$.
 	Since $\N_1\leq \widehat{\N}< \N^*$, there exists a circular partition $\widehat{\sigma}\in \C(\widehat{\N})$ and a circular partition $\sigma^*\in \C(\N^*)$ with $\sigma_0 \prec \widehat{\sigma} \preceq \sigma^*$. This implies 
 	that $\C(\N^*)$ contains two distinct circular partitions $\sigma$ and $\sigma^*$ such that $\sigma_0\preceq \sigma$ and $\sigma_0\preceq \sigma^*$;  
 		a contradiction in view of Lemma~\ref{lem:dom:compatible} and the fact that 
 	$\C(\N^*)$ is compatible. 
 	
 	 Finally, we consider the case that $\sigma$ corresponds to a cycle $C$ of at least four edges in $\N^*$ (i.e., $\widehat{\N}$ is obtained from $\N^*$ by  an edge contraction of $\N^*$ on 
 	 an edge of $C$ and $\sigma=\cpo(C)$).  %Then $\sigma$ is a proper circular partition.
 	  Let $C'$ be the cycle in $\widehat{N}$ obtained from
 	 $C$ by this edge-contraction and let $\sigma'=\cpo(C')$ denote the circular partition corresponding to $C'$. Then $(\C(\N^*)-\{\sigma\})\cup \{\sigma'\}=\C(\widehat{\N})$.  Now we 
 	 consider two possible subcases: either $\sigma\in W_0$ or $\sigma\in \lambda(F)$. 
 	
Assume first that $\sigma\in W_0$. 
%Consider the circular partition $\sigma_0 \in \C(\N_1)$ with $\sigma_0\preceq \sigma$.
Then we may further assume that $\sigma\in \C(\N_i)$, 
some $i\in\{1,2\}$, $i=1$, say.
	%  as a similar argument works for $\sigma\in \C(\N_2)$.
 %Then we have $\sigma_0=\sigma$ as otherwise $\sigma_0$ and $\sigma$ is incompatible, a contradiction. 
% Now the proof of this subcase is similar to that of the case of tiny cycle. That is, 
Since $\N_1\leq \widehat{\N}< \N^*$, there exists a circular partition $\widehat{\sigma}\in \C(\widehat{\N})$ and a circular partition $\sigma^*\in \C(\N^*)$ with $\sigma \prec \widehat{\sigma} \preceq \sigma^*$. This implies 
that $\C(\N^*)$ contains two circular partitions $\sigma$ and $\sigma^*$ with $\sigma \prec \sigma^*$,  
a contradiction to Lemma~\ref{lem:dom:compatible} and the fact that 
$\C(\N^*)$ is compatible. 

Finally, assume that  $\sigma\in  \lambda(F)$. Then there exist $\sigma_1\in
\C(\N_1)$ and $\sigma_2\in \C(\N_2)$ such that $\{\sigma_1,\sigma_2\}\in F$
and $\lambda(\{\sigma_1,\sigma_2\})=\sigma$. We claim that $\sigma_1\preceq \sigma'$, where $\sigma'=\cpo(C')\in \C(\widehat{\N})$. Assume for contradiction that this is
not the case. Then since $\C(\N_1)\dom \C(\widehat{\N})$ there exists a circular
partition $\sigma''\in \C(\widehat{\N})-\{\sigma'\}$ such that $\sigma_1\preceq \sigma''$.
Since $\sigma\not\in\C(\widehat{\N})$ and $\sigma''\in \C(\widehat{\N})-\{\sigma'\}\subseteq \C(\N^*)$,  it follows that $\sigma$ and $\sigma''$ are two distinct circular partitions in the compatible set
$\C(\N^*)$ such that $\sigma_1 \preceq \sigma$ and $\sigma_1\preceq \sigma''$ hold, 
% holds for two distinct circular partition $\sigma$ and $\sigma''$ in a compatible set of circular partition $\C(\N^*)$, which is impossible 
a contradiction in view of Lemma~\ref{lem:dom:compatible}.
Thus, $\sigma_1\preceq \sigma'$ as claimed.
%, we obtain $\sigma'=\sigma''$ which is impossible. 
Since a  similar argument also yields $\sigma_2\preceq \sigma'$  
it follows that 
$\sigma'$ is an upper bound for $\sigma_1$ and $\sigma_2$;
a contradiction to the fact that $\sigma'\prec \sigma$ and the assumption that 
$\sigma$ is a least upper bound for $\sigma_1$ and $\sigma_2$. This completes the proof of the theorem.
%we obtain the required contradiction, which completes the 
 \end{proof}

\section{Lower bounds} \label{sec:lower}

In this section, we investigate 
properties of greatest lower bounds of subsets in $(\pp(X),\le)$, which 
always exists since the trivial $X$-cactus is a lower bound for
any such subset. In particular, as a 
consequence of the main result of this
section (Theorem~\ref{thm:glb}), we characterize when a subset of $\pp(X)$ has 
the trivial $X$-cactus as a greatest lower bound (see Corollary~\ref{c:trivial}).

Greatest lower bounds in $\pp(X)$  are of interest since they can be considered as
``consensus networks''.  Indeed, in the poset of $X$-trees $(\mathcal T(X),\le)$,
the greatest lower bound for any subset $\mathcal T$ of $X$-trees is unique
and is known as the {\em strict-consensus tree} for $\mathcal T$ (cf. \cite{semple2003phylogenetics}).
%This fact also holds for subsets of $X$-trees in $\pp(X)$ (as can be seen 
%by applying the Theorem ~\ref{thm:glb} below).
However, for arbitrary sets of $X$-cactuses, greatest lower bounds
in $(\pp(X),\leq)$  are not necessarily unique. For example, 
for $X=\{1,2,3,4\}$ the $X$-cactuses $\N(\{[14|2|3]\})$ and $\N(\{[1|23|4]\})$
are both greatest lower bounds for the two $X$-cactuses $\N(\{[1|2|3|4]\})$ and $\N(\{[1|3|2|4]\})$.
The main result of this section (Theorem~\ref{thm:glb}) gives a 
characterization for when an $X$-cactus $\N$ is in the set $\glb(\N_1,\cdots,\N_m)$
consisting of all of the greatest lower bounds of a subset  $\{\N_1,\dots,\N_m\}\subseteq \pp(X)$.

%Given a set $\C$ of circular partitions $\sigma_1,\sigma_2,\cdots,\sigma_m$. Let 
%$\glb(\C)=\glb(\sigma_1,\cdots,\sigma_m)$ be the set of proper 
%circular partition $\sigma$ such that $\sigma$ is a greatest lower 
%bound of $\C$. If $\glb(\sigma_1,\cdots,\sigma_m)$ is non-empty, we 
%say these $m$ circular partitions have a {\em meet}. For example,  
%we have $\glb([1|2|3|4],[1|3|2|4)=\{[14|2|3],[1|23|4]\}$.

To state this result, we introduce some further terminology.
Suppose $\C_1,\cdots,\C_m$, some $m\geq 1$, are sets of circular partitions of $X$.
An element $(\mu_1,\cdots,\mu_m)$ in the product $\prod_{i=1}^m\C_i$ 
is said to have a {\em meet} if its set 
$\glb(\mu_1,\cdots,\mu_m)$ of greatest lower bounds under $\preceq$ is non-empty.  
A subset $\Gamma \subseteq \prod_{i=1}^m\C_i$ is called {\em feasible} if $\Gamma\not=\emptyset$, 
 for all distinct $(\mu_1,\cdots,\mu_m), (\mu'_1,\cdots,\mu'_m) \in \Gamma$ we have
$\mu_i\not=\mu'_i$, for all $1\le i \le m$, and every element in $\Gamma$
has a meet.  
A {\em meet realisation} of such a subset $\Gamma$ is a
subset $\C$ of $\C(X)$ that consists of precisely one meet  for 
each element in $\Gamma$, that is, $\C$ is the minimal subset of $\C(X)$ (under set inclusion) such that $|\glb(\mu_1,\cdots,\mu_m) \cap \C|=1$  holds for each element $(\mu_1,\cdots,\mu_m)$ in $\Gamma$. 
Note that this implies $|\C|\le |\Gamma|$.	
%Note that we use the convention that the emptyset is 
%the meet realisation of an empty feasible set.  

To illustrate these concepts, consider the sets $\C_p(\N_1)$ and $\C_p(\N_2)$ of 
proper circular partitions induced by the networks $\N_1$ an $\N_2$ 
considered in Figure~\ref{f:1_Fig_poset}. Then $\alpha=[1|2|3456789]$ and  
$ \beta_1=[678912|3|4|5]$ are two  proper circular partitions
in $\C_p(\N_1)$ and  $\beta_2=[678912|3|45]$ is a proper circular partition 
in $\C_p(\N_2)$. Since $\alpha\in \C_p(N_2)$,
 it follows that $\alpha$ is a meet for $(\alpha,\alpha)\in \C_p(\N_1)\times \C_p(\N_2)$.
 Furthermore, $\beta_2$ is a meet for $(\beta_1,\beta_2)\in \C_p(\N_1)\times\C_p(\N_2)$. 
 Thus, the set $\Gamma=\{(\alpha, \alpha), (\beta_1,\beta_2)\}\in \C_p(\N_1)\times \C_p(\N_2)$ is feasible. 
 For $\gamma=[7|89|123456]\in \C(\N_1)$ the set
 $\{(\gamma, \gamma)\}\cup \Gamma$ is a (maximal) feasible subset of 
 $\C_p(\N_1)\times \C(\N_2)$ and  
 $\C_p(\N_3)$ is a meet realization for it where $\N_3$ is the $X$-cactus in Figure~\ref{f:1_Fig_poset}. In fact, Theorem~\ref{thm:glb} below implies
 that $\N_3\in \glb(\N_1, \N_2)$.
 
%To illustrate these concepts assume that $\N_1,\ldots \N_m$,
%$m\geq 1$, is a set of $X$-cactuses and that $\C_i=\C(\T_i) $.
%Then $(\mu_1,\ldots,\mu_m)\in \prod_{i=1}^m C_i$ is a choice of
%partitions one from each set $\C_i$ and, provided
%a partition of $X$ exists
%that can be refined into $\mu_i$, $1\leq i\leq m$ then
%the finest such partition is the meet of $(\mu_1,\ldots,\mu_m)$.
%If $\Gamma$ is a selection of partitions one from each
%$\C_i$ such that no partition in some $\C_i$ is selected 
%more than once then $\Gamma$ is feasible if for
%each selection $\sigma$ of partitions in $\Gamma$ 
%there exists a partition $\pi_{\sigma}$ such that
%every partition in $\sigma$ is a refinement of $\pi_{\sigma}$.
%Clearly the partition $\pi_{sigma}$ might not be unique but 
%if it is for every selection $\sigma$ then the collection of
%of these unique meets is a meet realization of 
%$\Gamma$.

%For a set $\Sigma$ of phylogenetic $X$-trees the strict consensus
%tree $\mathcal T$ of the trees in $\Sigma$ is the phylogenetic $X$-tree that induces precisely the splits that are common to all
%trees in $\Sigma$. Furthermore, $\T=\glb(\Sigma)$.
%Extending this concept to $X$-nets by defining the strict consensus of a set $\Sigma$ of $X$-nets as the
%$X$-net $\N=\glb(\Sigma)$ then Lemma~\ref{lem:new} below implies
%that the set of splits induced by the strict consensus network
% is precisely the set of induced splits that are common
% to all $X$-nets in $\Sigma$.

Before stating Theorem~\ref{thm:glb}, we 
prove a useful lemma which describes how the splits
behave when taking greatest lower bounds in $(\pp(X), \leq)$. 

\begin{lem}\label{lem:new}
	Let $\{\N_1,\dots,\N_m\}\subseteq \pp(X)$, some $m\geq 1$.  If $\N \in \glb(\N_1,\dots,\N_m)$ then
	$\C_b(\N)=\bigcap_{i=1}^m \C_b(\N_i)$. 
	\end{lem}
	\begin{proof}
		%Note first that we may assume that $\C_b(\N)\not=\emptyset$ since otherwise the lemma clearly 			holds. So assume for the remainder that $\C_b(\N)\not=\emptyset$.
	Let $\C^*_b=\bigcap_{i=1}^m \C_b(\N_i)$. 
	%and  $\C^*_p=\prod_{i=1}^m\C_p(\N_i)$. 
	By Corollary~\ref{cor:bound:char}, we have $\C_b(\N)\subseteq \C_b^*$. 
	We may assume that $\C^*_b \not=\emptyset$ since otherwise the lemma clearly 			holds.
	
	To see that $\C^*_b \subseteq \C_b(\N)$, assume for contradiction that there exists a split $\sigma$ in $\C^*_b-\C_b(\N)$. 
	We claim that $\C(\N)\cup \{\sigma\}$ is compatible. 
	To this end, consider an arbitrary partition $\sigma'$  in $\C(\N)$. We need to show that $\sigma$ is compatible with $\sigma'$. 
	%Note that we have $\sigma'\not=\sigma$ and there are two subcases to consider. 
	Suppose first that $\sigma'\in \C_b(\N)$. Then  $\C_b(\N)\cup \{\sigma\}\subseteq \C_b(\N_1)$ implies that $\sigma$ and $\sigma'$ are both contained in $\C_b(\N_1)$. Since $\C_b(\N_1)$ is a compatible set of splits it follows that $\sigma$ and $\sigma'$ are compatible.
	 % either $\C_b(\N)=\emptyset$ or . 
	 Suppose next that $\sigma'\in \C_p(\N)$. As $\C_p(\N) \dom \C_p(\N_1)$ there exists a circular partition $\sigma_1$ in $\C_p(\N_1)$ 
	with  $\sigma'\preceq \sigma_1$. 
	Together with 	Lemma~\ref{lem:refinement:compatible} and the fact that $\sigma$ and $\sigma_1$ are two distinct compatible circular partitions in $\C(\N_1)$,  it follows that $\sigma$ and $\sigma'$ 	are compatible, which completes the proof the claim. 
	%that $\C(\N)\cup \{\sigma\}$ is compatible follows. 
	%Irrespective of whether or not $\C_p(\N)=\emptyset$, it follows that $\C(\N)\cup \{\sigma\}$ is 	compatible. 

	Since $\C(\N)\cup\{\sigma\}$ is a compatible set of circular partitions of $\C(X)$, it follows by 	Theorem~\ref{t:bijection.QX.CN} 
	that there exists an $X$-cactus $\N'$  with 
	$\C(\N')=\C(\N)\cup \{\sigma\}$. Since $\C(\N)\subset \C(\N')$, by Lemma~\ref{lem:dom:basic}(i) we have $\C(\N) \dom \C(\N')$ and hence $\N<\N'$ in view of Theorem~\ref{thm:poset:net:partition}
	because $\N\not=\N'$ as $\C(\N)\subset \C(\N')$. 
	%	, we obtain 	$\N < \N'$. 
	 For all $1\leq i\leq m$, since $\sigma $ is a split in $\C_b^*$ and $\C_b(\N)\subseteq \C^*_b$, it follows
	that $\C_b(\N')\subseteq \C_b^*\subseteq \C_b(\N_i)$ and that 
	$\C_p(\N')\dom \C_p(\N_i)$.
	By Corollary~\ref{cor:bound:char}, 
	$\N'$ is a lower bound for $\{\N_1,\dots,\N_m\}$ with $\N<\N'$; a contradiction
	to the assumption that $\N\in glb(\N_1,\ldots, \N_m)$.  
	\end{proof}

We now state and prove the main result of this section.

\begin{thm}\label{thm:glb}
	Let $\{\N_1,\dots,\N_m\}\subseteq \pp(X)$ and let $\N\in\pp(X)$.  Then
	$\N \in \glb(\N_1,\dots,\N_m)$ if and only if $\C_b(\N)=\bigcap_{i=1}^m \C_b(\N_i)$ and either
	(a) $\C_p(\N)=\emptyset$ and $\prod_{i=1}^m\C_p(\N_i)$ contains no feasible subset, or  
    (b) $\C_p(\N)\not=\emptyset$ and 
	$\C_p(\N)$ is a meet realization of some maximal feasible 
	subset of $\prod_{i=1}^m\C_p(\N_i)$ (under set inclusion).
\end{thm}

\begin{proof}
	Let $\C^*_b=\bigcap_{i=1}^m \C_b(\N_i)$ and  $\C^*_p=\prod_{i=1}^m\C_p(\N_i)$. 
	%It suffices to establish the equivalence of Statements~(i) and (ii) as the theorem's 
	%statement about trivial $X$-cactus is clearly a consequence.
	
	First, suppose $\N \in \glb(\N_1,\dots,\N_m)$. Then, by 
	Corollary~\ref{cor:bound:char}(ii), $\C_p(\N) \dom \C_p(\N_i)$ for $1\le i \le m$ and, 
	by Lemma~\ref{lem:new}, $\C_b(\N)=\C_b^*$.
	
	To see Statement~(a) suppose that $\C_p(\N)=\emptyset$, that is, $\N$ does not contain any cycles. 
	We need to show that $\C^*_p$ contains no feasible subset. 

	Suppose that this is not the case and that $\Gamma$ is a feasible 
	subset of $\C^*_p$. Then $\Gamma\not=\emptyset$, and for every
	$(\mu_1,\ldots, \mu_m)\in\Gamma$ we have that $\glb(\mu_1,\ldots, \mu_m)\not=\emptyset$. Now fix an element $(\mu_1,\ldots, \mu_m)\in\Gamma \subseteq\C_p^*$
	and a partition $\sigma\in \glb(\mu_1,\ldots,\mu_m)$.
	Since $\mu_1\in \C_p(X)$ and $\sigma\preceq \mu_1$, it follows that $\sigma$ must be a proper circular partition, that is, $\sigma\in \C_p(X)$.	
	By an argument similar to the one
		used in the proof of Lemma~\ref{lem:new} it follows that
	 $\C'=\C^*_b\cup\{\sigma\}$ is compatible. 
	 Therefore by Theorem~\ref{t:bijection.QX.CN}, there exists a $X$-cactus $\N'$ for which $\C(\N')= \C'$ 
	 holds. Since 
	$\C_b(\N)=\C_b^*\subset \C_b^*\cup\{\sigma\}=\C'$ it follows by 
	Lemma~\ref{lem:dom:basic}(i) and the assumption that $\C_p(\N)=\emptyset$ that $\C(\N)=\C_b(\N)\dom \C'=\C(\N')$. Hence,  $\N< \N'$ in view of  
	Theorem~\ref{thm:poset:net:partition}. 
	
	Next, we claim that 
	$\N'$ is a lower bound for $\{\N_1,\ldots, \N_m\}$.  
	%Since $\sigma\in\C_p(X)$ the definition of 		$\C'$ implies that $\C_p(\N')=\{\sigma\}$. 
	To this end, consider an arbitrary index $1\leq i\leq m$.	Since $\sigma\in\glb(\mu_1,\ldots,\mu_m)$, we have $\sigma\preceq \mu_i$ and hence 
		$\C_p(\N')=\{\sigma\}\dom \C(\N_i)$.
		Together with $\C_b(\N')=\C_b(\N)=\C^*_b\subseteq \C(\N_i)$ and  Corollary~\ref{cor:bound:char}(ii), it follows that $\N'\le \N_i$.
		Thus  $\N'$ is a lower bound for $\{\N_1,\ldots, \N_m\}$;
		a contradiction since $\N<\N'$ 
		 and $\N\in\glb(\N_1,\ldots,\N_m)$.
		Thus, $\C_p^*$ cannot 
	contain a feasible subset, which completes the proof of Statement~(a). 
		
	To see that Statement~(b) holds, suppose that $\C_p(\N) \neq \emptyset$. 
	Then $\C_p(\N) = \{\sigma_1,\dots,\sigma_k\}$ for some $k \ge 1$. 
	For each $1\le i \le k$, we construct an $m$-tuple $(\nu_{i,1}, \dots, \nu_{i,m})$ in $\C^*_p$ 
	by letting $\nu_{i,j}$, $1\le j \le m$,  be the circular partition   in $\C_p(\N_j)$ 
	with	$\sigma_i\preceq \nu_{i,j}$ (which exists because $\C_p(\N)\dom \C_p(\N_j)$). 
   We claim that $H=\{(\nu_{i,1}, \dots, \nu_{i,m}) \,:\, 1 \le i \le k\}$ is a feasible subset 
	of $\C^*_p$ and  that $\C_p(\N)$ is a meet realisation of $H$.  
	
	To see that the claim holds, note first that since
	$\C_p(\N)\dom \C_p(\N_j)$ for all $1\le j \le m$, it follows that
	$\nu_{i,j} \neq \nu_{l,j}$ for every pair $1\le i < l \le k$ 
	because a domination map is injective. To see that $\sigma_i\in \glb(\nu_{i,1}, \dots, \nu_{i,m})$  
	holds for all $1\le i \le k$, assume for contradiction that there exists 
	some $\widehat{\sigma}_i\in \glb(\nu_{i,1}, \dots, \nu_{i,m})$ with $\sigma_i\prec \widehat{\sigma}_i$. 
	Employing an argument similar to the one
	used in the proof of Lemma~\ref{lem:new} it follows that there exists an $X$-cactus $\widehat{\N}$ with $\C(\widehat{\N})=(\C(\N)-\{\sigma_i\})\cup \{\widehat{\sigma}_i\}$
	such that $\N<\widehat{\N}$ and $\widehat{\N}$ is a lower bound for 
	$\{\N_1,\dots,\N_m\}$; a contradiction since  $\N\in\glb(\N_1,\dots,\N_m)$. 
	Thus, $H$ is  feasible subset of $\C_p^*$.
	
	%	$\widehat{\C}=(\C(\N)-\{\sigma_i\})\cup \{\widehat{\sigma}_i\}$ is compatible, and hence for the $X$-cactus $\widehat{\N}$ with $\C(\widehat{\N}=\widehat{\C}$, we have 
	To see that  $\C_p(\N)$ is a meet realization of $H$, we need to show that $|\glb(\nu_{l,1},\ldots\nu_{l,m})\cap \C_p(\N)|=1$. To this end, note that
 for each pair $i,l$ with $1\le i \leq k$, $1\leq l \leq m$ and $i\not=l$, we have $\sigma_i\not \in \glb(\nu_{l,1}, \dots, \nu_{l,m})$ because otherwise  $\sigma_i \preceq \nu_{i,1}$ and $\sigma_i\preceq \nu_{l,1}$ both hold; a contradiction in view of  $\C_p(\N) \dom \C_p(\N_1)$ and Lemma~\ref{lem:dom:compatible}
 because $\nu_{i,l}\not=\nu_{l,1}$.
	Therefore $|\glb(\nu_{l,1},\ldots\nu_{l,m})\cap \C_p(\N)|=1$ and so $\C_p(\N)$ is a meet realisation of $H$. 
This completes the proof of the claim.
	
	It remains to show that $H$ is a maximal feasible subset of $\C_p^*$. 
	If not,  
	there exists an element $(\nu_{0,1}, \dots, \nu_{0,m}) \in \C^*_p-H$ 
	such that $H'=H\cup \{(\nu_{0,1}, \dots, \nu_{0,m})\}$ is a feasible subset of $\C_p^*$. 
	Let $\sigma_{0}\in \C_p(X)$ be a circular partition in $\glb(\nu_{0,1}, \dots, \nu_{0,m})$. Then we have $\sigma_{0}\not \in \C_p(\N)$. Indeed, assume for contradiction that $\sigma_{0}\in \C_p(\N)$. Then if $\sigma_0=\sigma_i$ for some $1\le i \le k$, then we have $\sigma_i\preceq \nu_{0,1}$ and $\sigma_i\preceq \nu_{0,i}$; a contradiction in view of $\C_p(\N) \dom \C_p(\N_1)$ and Lemma~\ref{lem:dom:compatible}
	because $\nu_{0,1}\not=\nu_{0,i}$. Thus, 
		$\sigma_{0}\not \in \C_p(\N)$.
	%	 by an argument similar to that in the last paragraph, and hence we have $\sigma_{0}\not \in \C(\N)$. 
	Furthermore, an argument similar to that in 	Lemma~\ref{lem:new} shows that  $\C(\N)\cup \{\sigma_{0}\}$ 
	is compatible. Hence there exists an $X$-cactus $\N'$ with 
	$\C(\N')=\C(\N)\cup \{\sigma_{0}\}$ in view of Theorem~\ref{t:bijection.QX.CN}.
	By Lemma~\ref{lem:dom:basic}(i) and Theorem~\ref{thm:poset:net:partition}, it follows 
	that $\N < \N'$ because $\sigma_{0}\not \in \C_p(\N)$. 
	
   Finally,  since $\sigma_{0}\in \C_p(X)$ we have
	$\C_b(\N')=\C_b(\N)$ and $\C_p(\N')=\{\sigma_{0}\}\cup\C_p(\N)=\{\sigma_{0},\sigma_1,\dots,\sigma_{k}\}$.
	Now for an arbitrary index $1\le j \le m$, consider the map $L_j: \C_p(\N') \to \C_p(\N_j)$ that maps each $\sigma_i$  to $\nu_{i,j}$, for all $0\leq i \le m$. Noting that $\sigma_i\preceq \nu_{i,j}$ for $0\le i \le k$, and $\nu_{i,j}\not =\nu_{l,j}$ for $0\le i < l\le k$ since $H'$ is a feasible subset of $\C_p^*$, it follows that $L_j$ is a domination map and hence $\C_p(\N') \dom \C_p(\N_j)$. 
	Together with $\C_b(\N')=\C_b(\N)=\C_b^*$,  
	Corollary~\ref{cor:bound:char} implies that $\N'$ is a lower 
	bound of $\{\N_1,\dots,\N_m\}$; a contradiction as $\N<\N'$ and $\N\in\glb(\N_1,\dots,\N_m)$. Thus $H$ is a maximal feasible subset of $\C^*_p$.

	We now show that the converse direction in the theorem holds.
	Assume for contradiction that $\N$ is such that the
	last statement in the theorem holds, but that $\N\not\in glb(\N_1,\ldots, \N_m)$. 
%Then, by assumption and Lemma~\ref{lem:new}, $\C_b(\N)=\C_b^*= \C_b(\N')$ must hold for every $X$-cactus $\N'$ such that $\N\leq \N'$ and $\N'\in\glb(\N_1,\ldots, \N_m)$.
 %Note that $\C_b(\N)=\C_b^*$ 
 We distinguish the cases that $\C_p(\N)=\emptyset$ 
 and that $\C_p(\N)\not=\emptyset$.
 
 First assume $\C_p(\N)=\emptyset$. Then, 	
 by Corollary~\ref{cor:bound:char}, $\N$ is a lower bound of $\{\N_1,\ldots, \N_m\}$. Since $\N\not\in glb(\N_1,\ldots, \N_m)$,
 there must exist an 
 $X$-cactus $\N'\in \glb(\N_1,\ldots, \N_m)$ with 
 $\N<\N'$. By Lemma~\ref{lem:new} and out assumption, we have $\C(\N')=\C^*_b=\C(\N)$. Hence 
 $ \C_p(\N')\not=\emptyset$ in view of $\N<\N'$.
Let  $\sigma$ be a circular partition in $\C_p(\N')$. Since $\C_p(\N') \dom \C_p(\N_j)$ and $\C_p(\N_j)$ is compatible, 
for every $1\le j \le m$, let $\mu_j\in\C_p(\N_j)$ denote
the necessarily unique circular partition  with $\sigma\preceq \mu_j$. Then $\mu=(\mu_1,\ldots, \mu_m)\in \C_p^*$. Furthermore, $\sigma\preceq\mu_i$ for all $1\leq i\leq m$ and  $\glb(\mu_1,\ldots, \mu_m)\not=\emptyset$ 
because $\sigma$ is a lower bound of $\{\mu_1,\ldots,\mu_m\}$ in $(\C(X),\preceq)$. Hence,
$\{\mu\}$ is a feasible subset of $\C_p^*$, a contradiction to Statement~(a). 
  
Now, assume $\C_p(\N)\not=\emptyset$
so that $\C_p(\N)=\{\sigma_1,\dots,\sigma_k\}$, some $k\ge 1$. 
Let $H$ be a maximal feasible subset of $\C^*_p$ such that  $\C_p(\N)$ is a meet realization of $H$ (which must exist by Statement~(b)).
%Let $k'\ge 1$ be the number of elements contained in $H$. 
Then the elements in $H$ can be enumerated as the $m$-tuples  $(\nu_{i,1},\dots, \nu_{i,m})$ for $1\le i \le k'=|H|$. 
For $1\le i <l \le k'$, we have $\glb(\nu_{i,1},\dots, \nu_{i,m}) \cap \glb(\nu_{l,1},\dots, \nu_{l,m})=\emptyset$ in view of Lemma~\ref{lem:dom:compatible} and the fact that $\nu_{i,1}\not=\nu_{l,1}$ are two distinct circular partitions contained $\C(\N_1)$ and $\C(\N_1)$ is compatible. 
As $\C_p(\N)$ is a meet realization of $H$,
$|\glb(\mu_1,\ldots,\mu_m)\cap \C_p(\N)|=1$, for all $(\mu_1,\ldots,\mu_m)\in H$,
and so it follows that $k'=k$. 
Swapping the indices if necessarily, we may assume that $\sigma_i \in \glb(\nu_{i,1},\dots, \nu_{i,m})$ holds for $1\le i \le k$.  
 Fix an arbitrary index $j\in\{1,\ldots, m\}$ and consider the map $L_j:\C(\N) \to \C(\N_j)$ that maps $\sigma_i$ to $\nu_{i,j}$ for $1\le i \le k$. Since $L_j$ is injective and $\sigma_i \preceq L_j(\sigma_i)=\nu_{i,j}$ holds for $1\le i \le k$, it follows that $L_j$ is a domination map. Hence $\C_p(\N) \dom \C_p(\N_j)$ for $1\le j \le m$. Together with  $\C_b(\N)=\C_b^*$ it follows 
 by Corollary~\ref{cor:bound:char} that $\N$ is a lower bound of $\{\N_1,\dots,\N_m\}$.

	We conclude the proof of the theorem by showing that  $\N\in glb(\N_1,\ldots, \N_m)$. 
	Suppose for contradiction that this is not the case.
	Then there exists an $X$-cactus $\N'\in \glb(\N_1,\dots,\N_m)$
	with $\N < \N'$.   
	Let $\C'_p=\C_p(\N')$. By 
	%Theorem~\ref{thm:poset:net:partition}, 
	Lemma~\ref{lem:new} and our assumption on $\N$ we obtain $\C_b(\N')=\C_b^*=\C_b(\N)$. It follows that
	$\C_p(\N)\dom\C_p(\N')$ and $\C_p(\N)\not=\C_p(\N')$. 
	By Lemma~\ref{lem:dom:compatible}, there exists a unique domination map $L:\C_p(\N)\to \C_p(\N')$. We consider the following two subcases.
	
	First, suppose that there exists some $\sigma\in \C_p(\N)$ with 
	$\sigma\prec L(\sigma)$, that is, $L(\sigma)\not=\sigma$. 	
	Without loss of generality, we may assume that
	$\sigma=\sigma_1$.
	By Lemma~\ref{lem:dom:basic}(ii) and $\C(\N')\dom \C(\N_j)$,
 there exists, for all $1\le j \le m$, a domination map $L_j:\C_p(\N')\to \C_p(\N_j)$. By Lemma~\ref{lem:dom:compatible} 
	it follows that $L_j(L(\sigma_1))=\nu_{1,j}$ for all $1\leq j\leq m$. 
	Hence, $L(\sigma_1)\preceq \nu_{1,j}$ for all such $j$.
	Thus, $L(\sigma_1)$ is a lower bound for $(\nu_{1,1},\ldots,\nu_{1,m})$; a contradiction 
	to the fact that $\sigma_1\in \glb(\sigma_{1,1},\ldots,\sigma_{1,m})$ and $\sigma_1\prec L(\sigma_1)$.

	Finally, suppose that $\sigma_i= L(\sigma_i)$ holds for all $1\le i \le k$.
	% $\sigma\in \C_p(\N)$. 
	Then we have $\C_p(\N)\subseteq \C_p(\N')$. Since $\C_p(\N)\not = \C_p(\N')$, 
	we may choose a circular partition $\sigma_0\in \C_p(\N')-\C_p(\N)$. 
	%Now an argument similar to the case  $\C_p(\N)=\emptyset$ shows that 
		 Since, for all $1\le j \le m$,  $\C_p(\N') \dom \C_p(\N_j)$ and $\C_p(\N_j)$ is compatible, there exists, by Lemma~\ref{lem:dom:compatible}, a unique domination map 
		  $L'_j:\C_p(\N')\to \C_p(\N_j)$ from $\C_p(\N')$ to $\C_p(\N_j)$. Put $\nu_{0,j}=L'_j(\sigma_0)$. 
		 % be the (necessarily unique) circular partition in $\C_p(\N_j)$ with $\sigma_0\preceq \nu_{0,j}$. Furthermore  		
		 Then $\nu_0=(\nu_{0,1},\ldots, \nu_{0,m})\in \C_p^*$ 
	and  $\glb(\nu_{0,1},\ldots, \nu_{0,m})\not=\emptyset$ because $\sigma_0$ is a lower bound of $\{\nu_{0,1},\ldots,\nu_{0,m}\}$ in $(\C(X),\preceq)$. 
	Furthermore, for each pair $1\le i \le k$ and $1\le j \le m$, we have $\nu_{0,j}=L'_j(\sigma_0)\not = L'_j(\sigma_i)=\nu_{i,j}$ as $L'_j$ is a domination map and therefore injective. 	
	Hence,
	$H\cup \{\mu\}$ is a feasible subset of $\C_p^*$; a contradiction to the assumption that $H$ is a maximal feasible subset of $\C_p^*$. 	
		This establishes the that  $\N\in glb(\N_1,\ldots, \N_m)$ and therefore completes the proof of  the case  $\C_p(\N)\not=\emptyset$.
		%from which the theorem follows.
% 
\end{proof}

Theorem~\ref{thm:glb} immediately implies

\begin{cor}\label{c:trivial}
	The trivial $X$-cactus is the greatest lower bound 
	for a set $\{\N_1,\dots,\N_m\}$ of $X$-cactuses if and only if $\bigcap_{i=1}^m \C_b(\N_i)=\emptyset$ and none of 
	the subsets in $\prod_{i=1}^m\C_p(\N_i)$ is feasible.
\end{cor}
%Note that in the case of a collection $\Sigma=\{\N_1,\ldots, \N_m\}$, $m\geq 1$, of phylogenetic 
%$X$-trees the sets $\prod_{i=1}^m\C_p(\N_i)$ and $\C_p(\N$)
%are empty as no phylogenetic $X$-tree can contain a cycle.
%Loosely speaking,  Theorem~\ref{thm:glb} then states that a 
%phylogenetic $X$-tree $\N$ can only
%be greatest lower bound  for $\Sigma$ (i.e. the strict consensus of the
%trees in $\Sigma$)
%if and only if the set of splits induced by $\N$ is the
%set of induced splits common to all $X$-trees in $\Sigma$.

\section{Discussion}\label{sec:discussion}

In this paper, we have introduced a new poset of $X$-cactuses and
shown that it has several interesting
structural properties. We conclude by listing 
some open problems and possible directions for future research.
 
\begin{itemize}
\item Is it possible to characterize upper bounds for sets of
$X$-cactuses, for example, generalizing Theorem~\ref{thm:up}? Also, 
it is known that a collection of $X$-trees has an 
upper bound in $(\mathcal T(X),\le)$ if and only if
every pair of trees in the collection does \cite[Theorem 3.3.3]{semple2003phylogenetics}.
Is this also true for general collections of $X$-cactuses in $(\pp(X),\le)$?

\item As mentioned above, lower bounds for
collections of $X$-cactuses in $(\pp(X),\le)$ are
of interest as they could be used as consensus networks.
Bearing this in mind, is it possible to find an efficient algorithm to compute a %(or all) 
greatest lower bound for a set of $X$-cactuses?
Our results in Section~\ref{sec:lower} provide some insights into this problem, 
however, the computational complexity of this problem remains unresolved.

\item It could be of interest to further study structural
properties of $(\pp(X),\le)$. For example, what 
are properties of the M\"obius function of this poset? Also,
are there alternative ways to define partial orderings of $\pp(X)$?

\item Can encodings and
partial orders be defined for other classes of phylogenetic networks, such as 
``level-$k$'' networks or ``explicit'' networks (see \cite[Chapter 10]{steel2016phylogeny} for 
a recent overview of phylogenetic networks and definitions of these terms).

\item Finally, note that the partially ordered set $(\mathcal T(X),\le)$ of 
$X$-trees is intimately related to certain complexes and spaces 
of phylogenetic trees  \cite{ardila2006bergman,trappmann1998shellability}.
It would be of interest to understand how the structure of the
so-called order complex of $(\pp(X),\le)$ might be related to 
phylogenetic network spaces such as those described in, for
example, \cite{huber2016transforming} and \cite{devadoss2019split}.
\end{itemize}

\subsubsection*{Acknowledgement}
Francis, Huber and Moulton thank the Royal Society for its support. Huber and
Moulton also thank Western Sydney University for its hospitality.

\bibliographystyle{plain}
\bibliography{Xnets}
		
\end{document}